# Dispersion Engineered Frequency Tunable Delay Platform based on Magnetostatic Surface Waves

Chin-Yu Chang[1], Xingyu Du[1], Shun Yao[1], Tao Wang[1], Shuxian Wu[1], and Roy H. Olsson III[1]

[1]Department of Electrical and System Engineering, University of Pennsylvania, PA, USA

*Correspondence: Roy H. Olsson III (rolsson@seas.upenn.edu)

## Abstract

Reconfigurable radio frequency front ends in modern radar and wireless systems require delay elements that simultaneously offer low-loss, low noise, compact form factor, and wideband frequency agility. Yet electromagnetic, acoustic, photonic, and active-circuit delay technologies each fail to deliver this combination. Here we report a microwave delay platform based on magnetostatic surface waves (MSSWs) in microfabricated 18 μm yttrium iron garnet (YIG) waveguides. By co-engineering the spin wave dispersion with the radiation impedance of meander-line transducers grants pitch-controlled access to distinct dispersive or near-constant group delay regimes. Continuously tuned from 6 to 19.6 GHz under magnetic bias, the delay lines deliver group delays of 3.3 to 42.8 ns at insertion losses of 2.5 to 10.1 dB and nonreciprocal isolation of 24 to 39 dB, all measured directly into 50 Ω without external impedance matching. Length-resolved characterization yields unit-time propagation losses of 56 to 109 dB·μs$^{-1}$ and propagation $Q$-factors that rise monotonically from 3002 to 4893 across the operating range, exceeding state-of-the-art fixed frequency acoustic delay lines at every benchmarked frequency. These results establish microfabricated YIG as a versatile, low-loss microwave platform for next-generation reconfigurable RF signal processing.

Modern radar and wireless communication systems increasingly rely on real time control of microwave signals across multiple frequency bands. At the center of this control problem are delay lines. They set the timing and phase evolution of radio frequency (RF) waveforms and support two distinct signal processing regimes. In true-time-delay operation, nearly frequency-independent group delay preserves wideband beamforming accuracy by suppressing beam squint effects in phased array antennas, as shown in Fig. 1b [1], [2]. In dispersive signal processing, frequency-dependent group delay can be used for pulse generation/compression in advanced radar systems and chirped signal processing, as shown in Fig. 1c [3]. Although these regimes place different demands on group delay dispersion, next-generation delay elements would ideally combine low-loss, high delay density, and frequency-agile operation. Frequency-agile operation is especially important for reconfigurable RF front ends, but current delay line technologies are largely constrained to fixed operating frequency bands. As RF systems move toward multiband operation, dynamic spectrum access, and adaptation to congested electromagnetic environments, fixed-frequency delay elements often require multiple dedicated signal paths or front end modules. This hardware duplication increases system complexity and becomes especially problematic in unmanned aerial vehicles, low-Earth-orbit satellites, and mobile platforms, where compact and reconfigurable hardware is required under stringent size, weight, power, and cost constraints (SWaP-C) [4]. These challenges motivate the development of compact, low-loss, and frequency tunable delay elements for next-generation RF front ends.

Existing delay line technologies face different tradeoffs among achievable insertion loss, delay density, propagation loss, and system complexity. Electromagnetic transmission lines are simple and broadband, but their high group velocity yields a low delay density and therefore requires long physical path lengths to obtain nanosecond-scale delays. Acoustic delay lines (ADL) based on surface or plate acoustic waves provide much higher delay density because of their much lower group velocity. However, the operational frequency of ADLs depends on the pitch between interdigital transducers, requiring meticulous lithography process that becomes increasingly demanding at higher frequencies. Moreover, the propagation loss of acoustic waves increases rapidly as the frequency is increased due to the inherent $f{\cdot}Q$ product limitation [5], [6]. CMOS delay circuits offer electronic reconfigurability and active gain, yet they typically provide limited delay density while requiring substantial power consumption and introducing higher noise than passive delay elements [7], [8], [9], [10]. Photonic delay lines can provide broad RF bandwidth and nearly frequency independent delay time over a wide band by modulating electrical signals onto optical carriers. However, practical implementation in RF systems requires optical sources, electro-optic modulators, photodetectors, and supporting active circuits. These interfaces increase system complexity, power consumption, and noise, while the high optical group velocity still makes low-loss delay accumulation challenging when evaluated per unit time [11], [12], [13], [14], [15], [16], [17], [18], [19], [20]. Together, these approaches illustrate a persistent trade off among delay density, insertion loss, frequency coverage, and system complexity, highlighting the difficulty of realizing a

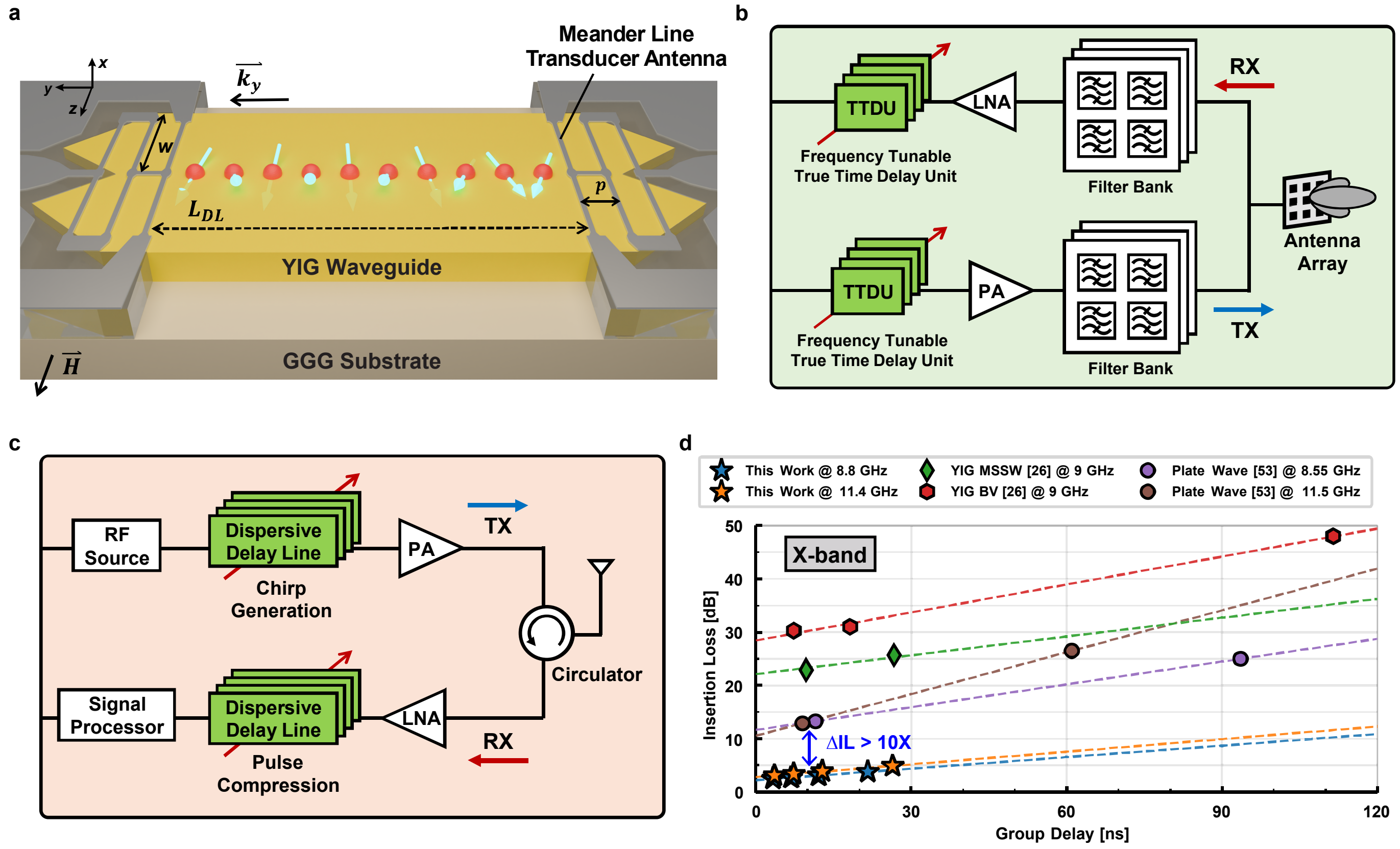

Fig. 1 (a) 3D illustration of the proposed MSSW YIG delay line with meander-line transducer antenna (b) Simplified schematic of a phased array-based radio frequency front end module with true time delay units. (c) Simplified schematic of a radar system with a dispersive delay line for chirp generation and pulse compression. (d) Benchmark of insertion loss versus group delay for the measured MSSW YIG delay lines and state-of-the-art YIG and acoustic delay lines operating at X-band.

compact, low-loss, frequency tunable delay platform for RF signal processing.

Magnetostatic spin waves in yttrium iron garnet (YIG) have emerged as an attractive platform for frequency tunable RF signal processing devices, offering magnetic field frequency tunability, low-loss propagation, relatively slow group velocity, and intrinsic nonreciprocity [21], [22], [23], [24], [25], [26], [27], [28]. They have been widely used for tunable filter synthesis based on forward-volume, backward-volume, and surface-wave modes [22], [23], [24], [25], [28], [29], [30]. These same attributes make magnetostatic spin waves attractive for frequency tunable delay elements. Magnetic bias can provide frequency agility, while slow spin wave propagation can support compact delay accumulation [31]. Prior YIG-based magnetostatic-wave delay lines have explored backward volume (BV) [32], magnetostatic surface wave (MSSW) [26], [33], [34], [35], and hybrid forward volume-BV [36] implementations. These studies show that YIG can provide long delay and magnetic field tunability. At the same time, their delay responses are generally governed by the strong dispersion of the underlying magnetostatic-wave modes. This makes YIG delay lines naturally relevant to dispersive delay applications, while also raising the question of how dispersion, loss, and delay accumulation should be managed for near-constant-delay operation. For microfabricated YIG waveguides, this propagation behavior remains less systematically studied. Developing MSSW delay elements therefore provides a route toward compact frequency-agile delay lines while establishing microfabricated YIG as a low-loss microwave propagation platform.

In this study, we demonstrate a frequency tunable delay line based on magnetostatic surface waves (MSSWs) in microfabricated YIG waveguides. By co-engineering the spin wave dispersion and the radiation impedance of meander-line transducers, we identify distinct delay response regimes. The low spurious pitch design provides low-loss delay accumulation with reduced spurious mode excitation. It operates from 6 to 19.6 GHz and exhibits delay times of 3.3 to 42.8 ns, insertion losses of 2.5 to 10.1 dB, and minimum in-band isolation of 24 to 39 dB. Measurements across multiple waveguide lengths further reveal low unit-time propagation loss of 56-109 dB·μs$^{-1}$, with propagation $Q$-factors ($Q_{PL}$) from 3002 to 4893 across 6-19.6 GHz. The MSSW delay line therefore advances well beyond state-of-the-art magnetostatic spin wave and acoustic wave delay line performance by combining this low propagation loss with substantially lower absolute insertion loss at comparable group delays, as benchmarked for X-band operation (8-12 GHz) in Fig. 1d. An alternative flatter delay pitch design produces a more uniform group delay response while showing stronger spurious mode excitation. This response indicates a pathway toward quasi-constant-delay operation versus frequency and motivates further suppression of spurious modes. Together, these measurements establish transducer-selected MSSW propagation as a design principle for microwave delay lines. The platform provides low-loss, frequency-agile delay accumulation and offers a route to engineering both dispersive-delay functionality and quasi-constant-delay operation.

# Results

## Dispersion Analysis and Transducer Design

Magnetostatic surface waves, also known as Damon-Eshbach modes, are dipole-dominated spin wave modes supported by a tangentially magnetized YIG film when the wavevector is perpendicular to the static magnetization, as shown in Fig. 1a. Their propagation behavior is governed by the dispersion relation given by Eq. (1)-(3)

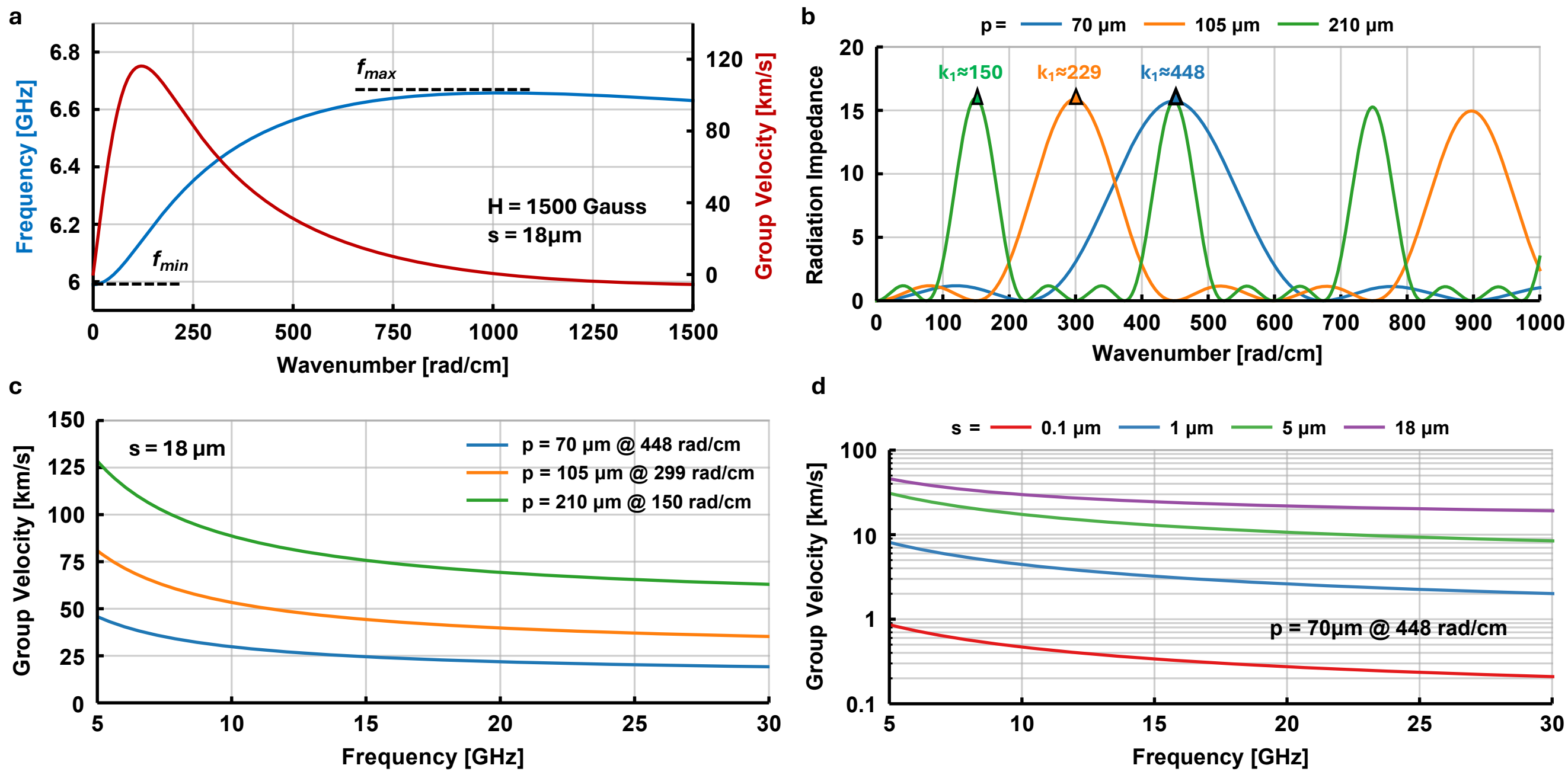

Fig. 2 (a) Dispersion curve of MSSW with 18 μm of YIG and magnetic field 1500 Gauss. (b) Radiation impedance of the meander-line transducer with different pitches. (c) Group velocity of the MSSW with fixed YIG thickness of 18 μm when excited at different wavenumbers by varying the pitch of the meander-line transducers. (d) Group velocity of the MSSW excited at a wavenumber of 448 rad/cm with different YIG thickness.

$$f(k_y)^2 = \gamma^2[H + M_s(1 - P + \alpha k^2)]\left[H + M_s(P\frac{k_y^2}{k^2} + \alpha k^2)\right] \quad (1)$$

$$P = 1 - \frac{1 - e^{-ks}}{ks} \quad (2)$$

$$k^2 = k_z^2 + k_y^2 \quad (3)$$

where $H$ is the applied magnetic bias field, $M_s$ is the saturation magnetization, $\gamma$ is the gyromagnetic ratio, $k_y$ is the propagation wavenumber, $k_z$ is the transverse wavenumber, $\alpha$ is the exchange stiffness, and $s$ is the YIG thickness [37], [38]. In the present finite width waveguide, the transverse wavenumber $k_z$ is quantized by the waveguide edges. Under the magnetic-wall boundary condition, $k_z$ is given by $k_z = n\pi/w$, where $n$ is the width-mode order and $w$ is the waveguide width. In contrast, $k_y$ is the longitudinal propagation wavenumber along the waveguide and determines how the excited MSSW propagates. Therefore, for a given magnetic field and waveguide geometry, Eq. (1)-(3) define a $k_y$-dependent MSSW dispersion with a finite allowable propagation band bounded by $f_{\min}$ and $f_{\max}$. Fig. 2a shows the calculated dispersion curve at 1500 Gauss magnetic bias for $w$ = 200 μm and 18 μm thick YIG waveguide. The dispersion indicates that, at a fixed magnetic field, MSSW propagation is allowed only within the frequency range between $f_{\min}$ and $f_{\max}$. This allowable propagation band directly sets the usable frequency window of the delay line and narrows as the magnetic bias field increases, as detailed in supplementary note 1. The corresponding group velocity is also extracted from the slope of the dispersion curve, $v_g = \partial\omega / \partial k_y$. As shown in Fig. 2a, the group velocity varies strongly with wavenumber and gradually decreases as the excited MSSW approaches the upper band edge. This unique behavior also leads to progressively less efficient wave transport near the band edge and contributes to a sharp spectral cutoff outside the usable MSSW band [22].

While the MSSW dispersion sets the allowable propagation band, the transducer determines which part of that dispersion is excited and measured experimentally. A meander-line transducer is used here because its periodic current pattern provides stronger wavenumber selectivity and more efficient coupling to the MSSW. This selectivity is described by the radiation impedance profile of the transducer, given by Eq. (4)

$$T_m(k_y) = \frac{R_m}{R_0 l} = \left[\frac{\sin(W_T k_y/2)}{W_T k_y/2}\right]^2 \left[\frac{\sin(pNk_y/2)}{\cos(pk_y/2)}\right]^2 \tag{4}$$

where $T_m$ is the normalized radiation impedance, $R_m$ is the radiation impedance, $R_0$ is a material and pitch dependent constant, $l$ is the transducer arm length, $W_T$ is the transducer width, $p$ is the meander pitch, and $N$ is the number of meander periods [39]. The first term defines the spectral envelope associated with the finite transducer width, whereas the second term represents the array factor introduced by the periodic meander geometry. Together, these terms determine how efficiently the transducer couples to MSSW modes at different wavenumbers. Fig. 2b shows the calculated normalized radiation impedance profile with different transducer pitches. The different transducer pitches preferentially excite MSSWs at different longitudinal wavenumbers. Taking the first peak as a reference, pitches of 70, 105 and 210 μm correspond to approximately 448, 299 and 150 rad·cm$^{-1}$, respectively. The transducer pitch therefore directly determines which $k_y$ region of the MSSW dispersion is sampled experimentally, and thus which group velocity regime is accessed.

As a delay-element platform, MSSWs offer a substantial design space rooted in their distinctive dispersion behavior. Since the applied magnetic field sets the operating frequency, the pitch-selected group velocity at each wavenumber can be traced continuously across the magnetically tuned frequency range. Fig. 2c shows that different transducer pitches access markedly different group-velocity regimes, and that this distinction persists over the full 5 to 30 GHz span. A further degree of freedom is provided by the YIG thickness. As shown in Fig. 2d, the group velocity changes strongly with film thickness, spanning more than

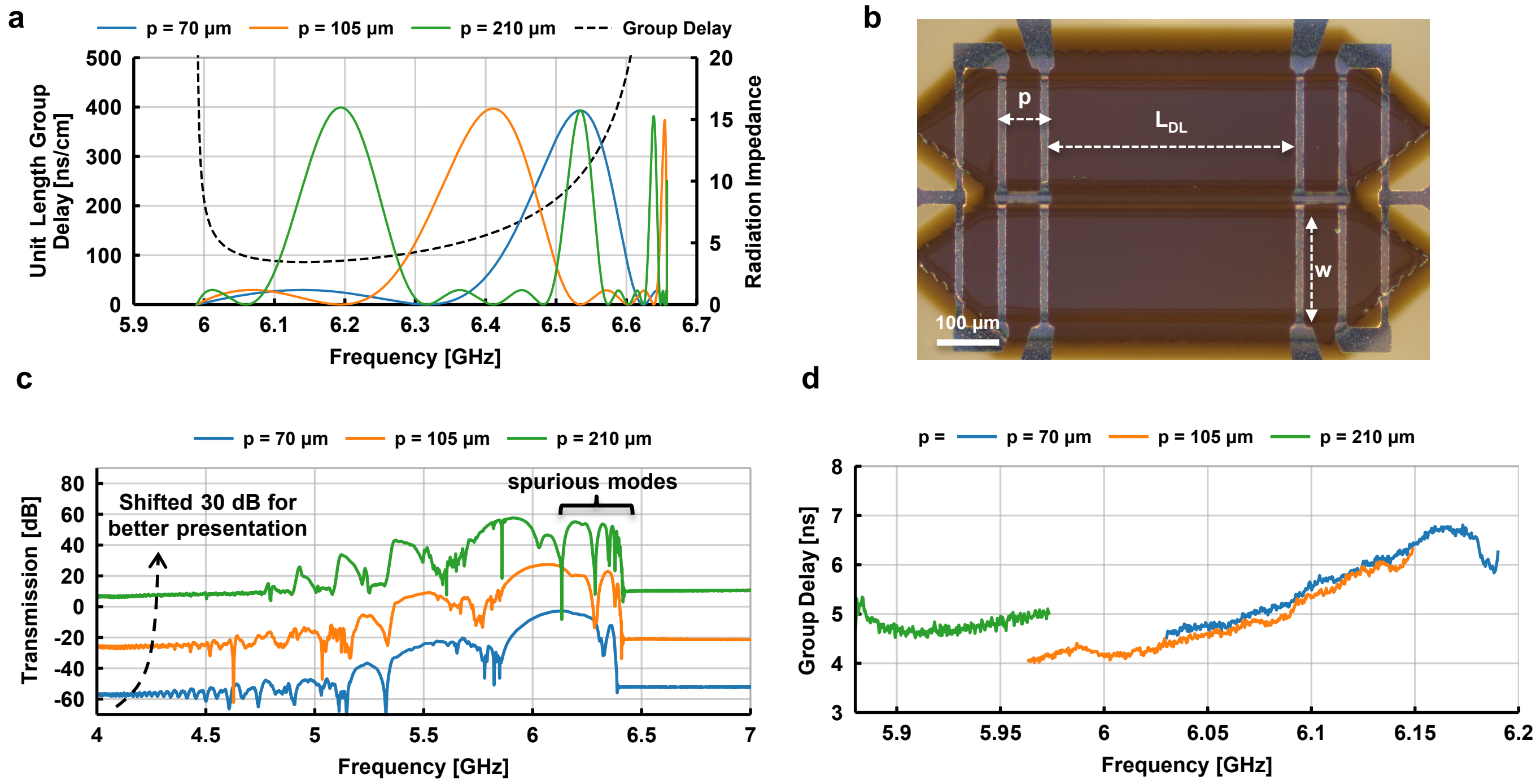

Fig. 3 (a) Calculated unit length group delay overlay with the calculated radiation impedance versus frequency. (b) Optical image of the fabricated MSSW YIG delay line. (c)(d) Measured delay line frequency response and group delay with meander-line transducer pitches of 70, 105, and 210 µm at 1500 Gauss applied magnetic field.

two orders of magnitude from sub-km·s$^{-1}$ for 100 nm YIG to several tens of km·s$^{-1}$ for 18 µm YIG. These results underscore the versatility of MSSW waveguides as a delay element platform. Their propagation characteristics can be tuned over a wide range by varying both the transducer geometry and the waveguide dimensions.

The design implications of dispersion and radiation impedance become clearer when recast in the frequency domain. Fig. 3a plots the unit-length group delay as a function of frequency. The bowl-shaped profile reflects the bounded propagation band of MSSWs at a fixed magnetic field. The radiation impedance can also be mapped onto the frequency domain through the dispersion curve. This representation reproduces the pitch-dependent group velocity variation identified in Fig. 2c and reveals where MSSW modes are excited in the spectrum. It also shows how many excitation points appear within the accessible band. These features govern both spurious mode excitation and the flatness of the group delay response for different transducer pitches. Fig. 3b shows an optical image of the proposed YIG MSSW delay line after fabrication. The design concept of the proposed devices follows our prior work [30], adopting a two-cavity meander-line geometry to improve impedance matching, together with triangular waveguide terminations to reduce internal MSSW circulation and standing-wave-related spurious responses. The detailed fabrication process and measurement setup are presented in the Methods. Fig. 3c and d show the measured frequency response and extracted in-band group delay for different transducer pitch designs. The results reveal two distinct delay response regimes. The $p$ = 70 μm design provides a cleaner spectrum with reduced spurious mode excitation, whereas the $p$ = 210 μm design produces a flatter in-band group delay response with stronger spurious mode excitation. This comparison shows that transducer pitch design can tailor the MSSW delay response toward low-spurious dispersive operation or quasi-constant-delay operation.

## Frequency Tunable MSSW Delay Lines

Based on the dispersion and radiation impedance analysis presented above, two meander-line pitch designs were fabricated to demonstrate pitch-dependent MSSW delay responses. Four $p$ = 70 μm devices with $L_{\mathrm{DL}}$ = 70, 210, 420, and 840 μm and two $p$ = 210 μm devices with $L_{\mathrm{DL}}$ = 210 and 840 μm were fabricated to evaluate frequency tunability, length-dependent delay accumulation, and propagation loss across different pitch designs. Fig. 4 shows the measured $S_{12}$ transmission responses of these devices under magnetic bias fields from 1500 to 6500 Gauss. Across the measured designs, the MSSW passband shifts from 6 to 19.6 GHz as the magnetic field increases. This confirms frequency tunable delay line operation across different transducer geometries and waveguide lengths. The measured spectra also preserve the pitch-dependent characteristics predicted in Fig. 3. The $p$ = 70 μm devices show a cleaner dominant passband, whereas the $p$ = 210 μm devices show stronger spurious mode excitation. This length scaled device set separates propagation dependent delay and loss from transducer coupling, enabling direct extraction of accumulated group delay and propagation loss.

Fig. 5a and b provide representative examples for comparing the group delay responses of the two pitch designs at 3500 Gauss. The $p$ = 70 μm design shows a dispersive delay response, with the center frequency group delay increasing from 3.6 ns for $L_{\mathrm{DL}}$ = 70 μm to 26.4 ns for $L_{\mathrm{DL}}$ = 840 μm. In comparison, the $p$ = 210 μm design shows a more uniform in-band group delay, with averaged in-band delays of 5.3 ns and 12.5 ns for $L_{\mathrm{DL}}$ = 210 and 840 μm, respectively. This comparison shows that transducer pitch can tailor the MSSW delay response from dispersive delay operation toward quasi-constant-delay operation. Across the 6-19.6 GHz tuning range, the $p$ = 70 μm devices exhibit center frequency delay times from 3.3 to 42.8 ns, IL from 2.5 to 10.1 dB, unit-time propagation losses of 55.7-109 dB·μs$^{-1}$, and $Q_{\mathrm{PL}}$ from 3002 to 4893. The $p$ = 210 μm devices exhibit delay times from 4 to 14 ns and IL from 2.3 to 11 dB, corresponding to extracted unit-time propagation losses of 221-655 dB·μs$^{-1}$ and $Q_{\mathrm{PL}}$values from 727 to 850. These results indicate two distinct pitch selected delay regimes. The $p$ = 70 μm design provides a dispersive delay response with extremely low IL, low propagation loss, and reduced spurious mode excitation, whereas the $p$ = 210 μm design provides a flatter group delay response with low IL but larger extracted propagation loss. The flatter delay response of the $p$ = 210 μm design allows nearly the same delay to be maintained across the operating band, suggesting a pathway toward quasi-constant-delay operation after further spurious mode suppression. Zoomed in transmission spectra and in-band group delay responses for all devices at each magnetic bias field are provided in supplementary note 2.

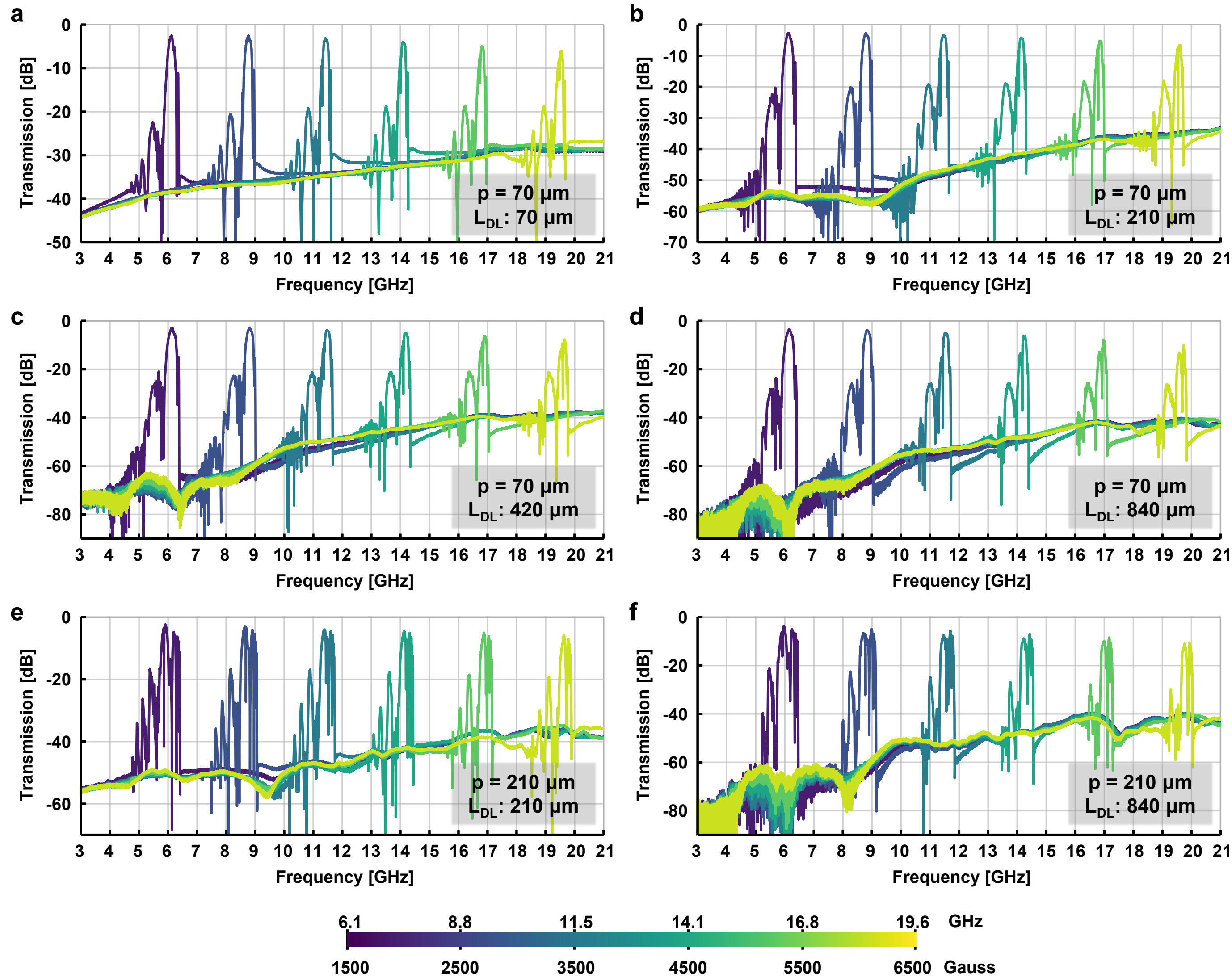

Fig. 4 Measured $S_{12}$ transmission spectra of MSSW YIG delay lines under magnetic bias fields from 1500 to 6500 Gauss. (a) - (d) devices with $p$ = 70 μm and $L_{DL}$ = 70, 210, 420, and 840 μm respectively. (e) and (f) Devices with $p$ = 210 μm and $L_{DL}$ = 210 and 840 μm, respectively. The passband is continuously tuned from 6 to 19.6 GHz as the magnetic bias field increases.

Fig. 5c further compares the proposed devices with representative YIG based delay lines using different magnetostatic wave modes, including backward volume (BV) [32], magnetostatic surface wave (MSSW) [26], [33], [34], [35], and hybrid forward volume-BV [36] operation. These prior demonstrations span YIG

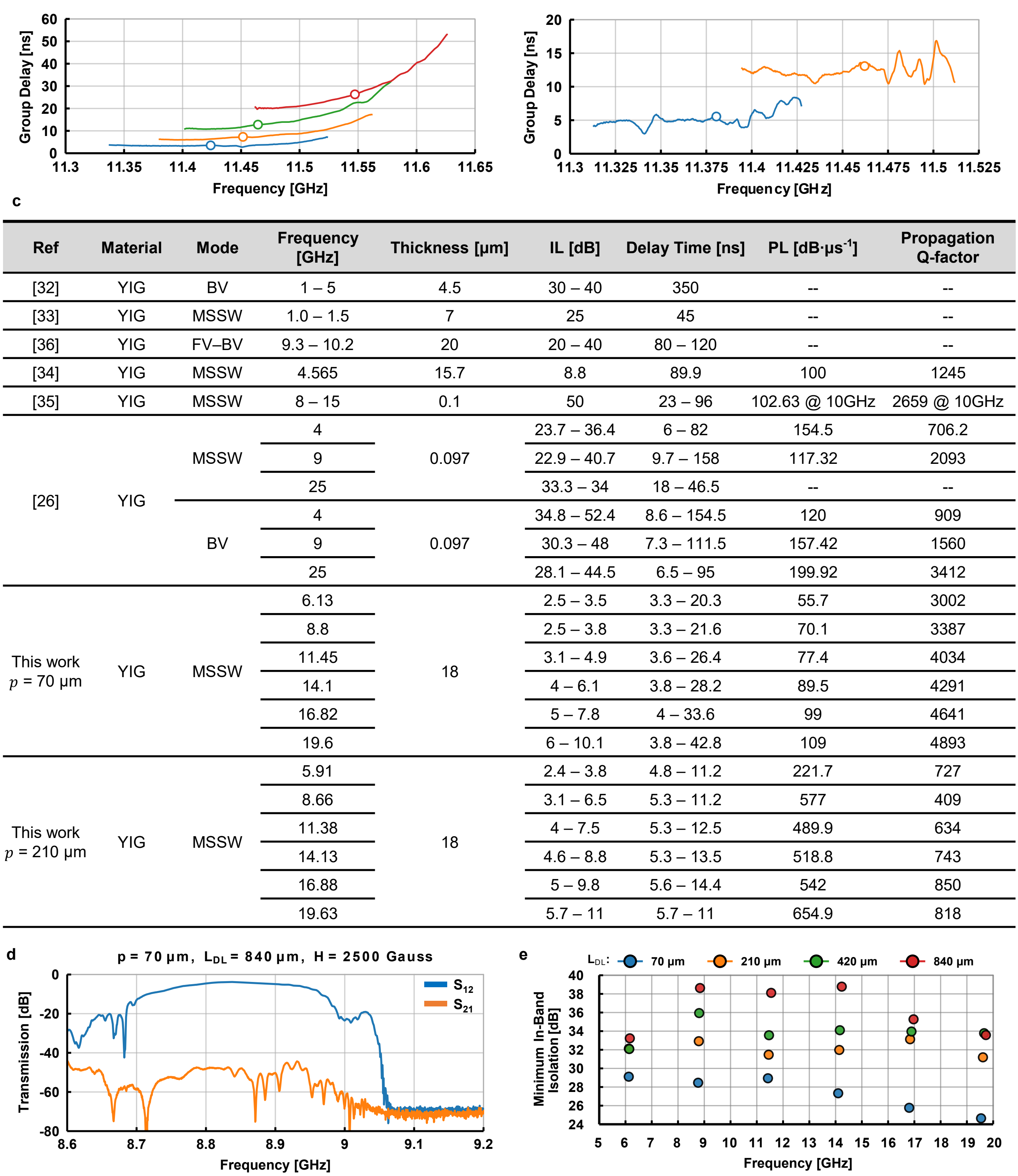

| Ref | Material | Mode | Frequency [GHz] | Thickness [μm] | IL [dB] | Delay Time [ns] | PL [dB·μs$^{-1}$] | Propagation Q-factor |
|---|---|---|---|---|---|---|---|---|
| [32] | YIG | BV | 1 – 5 | 4.5 | 30 – 40 | 350 | -- | -- |
| [33] | YIG | MSSW | 1.0 – 1.5 | 7 | 25 | 45 | -- | -- |
| [36] | YIG | FV–BV | 9.3 – 10.2 | 20 | 20 – 40 | 80 – 120 | -- | -- |
| [34] | YIG | MSSW | 4.565 | 15.7 | 8.8 | 89.9 | 100 | 1245 |
| [35] | YIG | MSSW | 8 – 15 | 0.1 | 50 | 23 – 96 | 102.63 @ 10GHz | 2659 @ 10GHz |
| [26] | YIG | MSSW | 4 | 0.097 | 23.7 – 36.4 | 6 – 82 | 154.5 | 706.2 |
| | | | 9 | | 22.9 – 40.7 | 9.7 – 158 | 117.32 | 2093 |
| | | | 25 | | 33.3 – 34 | 18 – 46.5 | -- | -- |
| | | BV | 4 | 0.097 | 34.8 – 52.4 | 8.6 – 154.5 | 120 | 909 |
| | | | 9 | | 30.3 – 48 | 7.3 – 111.5 | 157.42 | 1560 |
| | | | 25 | | 28.1 – 44.5 | 6.5 – 95 | 199.92 | 3412 |
| This work $p$ = 70 μm | YIG | MSSW | 6.13 | 18 | 2.5 – 3.5 | 3.3 – 20.3 | 55.7 | 3002 |
| | | | 8.8 | | 2.5 – 3.8 | 3.3 – 21.6 | 70.1 | 3387 |
| | | | 11.45 | | 3.1 – 4.9 | 3.6 – 26.4 | 77.4 | 4034 |
| | | | 14.1 | | 4 – 6.1 | 3.8 – 28.2 | 89.5 | 4291 |
| | | | 16.82 | | 5 – 7.8 | 4 – 33.6 | 99 | 4641 |
| | | | 19.6 | | 6 – 10.1 | 3.8 – 42.8 | 109 | 4893 |
| This work $p$ = 210 μm | YIG | MSSW | 5.91 | 18 | 2.4 – 3.8 | 4.8 – 11.2 | 221.7 | 727 |
| | | | 8.66 | | 3.1 – 6.5 | 5.3 – 11.2 | 577 | 409 |
| | | | 11.38 | | 4 – 7.5 | 5.3 – 12.5 | 489.9 | 634 |
| | | | 14.13 | | 4.6 – 8.8 | 5.3 – 13.5 | 518.8 | 743 |
| | | | 16.88 | | 5 – 9.8 | 5.6 – 14.4 | 542 | 850 |
| | | | 19.63 | | 5.7 – 11 | 5.7 – 11 | 654.9 | 818 |

Fig. 5 (a) Extracted in-band group delay of the $p$ = 70 μm devices at 3500 Gauss for $L_{DL}$ = 70, 210, 420, and 840 μm. (b) Extracted in-band group delay of the $p$ = 210 μm devices at 3500 Gauss for $L_{DL}$ = 210 and 840 μm. (c) Performance comparison between the proposed MSSW YIG delay lines and previously reported YIG delay lines based on different magnetostatic wave modes. (d) Measured $S_{12}$ and $S_{21}$ transmission spectra of the $p$ = 70 μm, $L_{DL}$ = 840 μm device at 2500 Gauss, showing strong nonreciprocity. (e) Minimum in-band isolation of the $p$ = 70 μm devices across different operating frequencies and delay line lengths.

thicknesses from the sub-micrometer regime to tens of micrometers and use different transducer schemes, including straight line, microstrip, and coplanar waveguide excitation. Sub-micrometer YIG waveguides can provide high delay density because of their reduced group velocity, but they generally show larger insertion loss and higher unit-time propagation loss. This behavior is consistent with the stronger influence of interface imperfections, magnetic inhomogeneity, and reduced volume-to-interface ratio in thinner YIG films [40]. In comparison, the 18 μm thick MSSW delay lines demonstrated here provide a favorable balance between delay density, absolute insertion loss, and propagation loss. The $p$ = 70 μm design serves as the low-loss benchmark, combining low IL, low unit-time propagation loss, high $Q_{\mathrm{PL}}$, and reduced spurious mode excitation over the 6.0-19.6 GHz tuning range. The $p$ = 210 μm design instead represents a complementary pitch selected regime that preserves low absolute IL and provides a flatter group delay response, although its effective propagation loss is higher. This comparison positions the microfabricated YIG MSSW platform as a strong candidate for low-loss, high frequency, and frequency tunable RF delay elements, with transducer pitch providing an additional route to tailor the delay response.

In addition to low-loss delay accumulation, the devices retain the intrinsic nonreciprocity of MSSW propagation. This directionality is advantageous for RF delay line applications. In conventional reciprocal delay lines, e.g., acoustic delay lines, reflections between input and output transducers can generate triple-transit echoes. In contrast, nonreciprocal MSSW propagation does not support the reverse spin wave path required for such round trip echo formation, thereby preserving a cleaner forward propagating delayed response. Fig. 5d shows representative $S_{12}$and $S_{21}$responses of the $p$ = 70 μm and $L_{\mathrm{DL}}$ = 840 μm device at 2500 Gauss, where forward MSSW transmission is accompanied by strong reverse isolation. Fig. 5e summarizes the minimum in-band isolation across operating frequencies and waveguide lengths. The devices maintain 24-39 dB of in-band isolation across the tuning range, establishing the microfabricated YIG MSSW waveguide as a low-loss, frequency-agile, and nonreciprocal delay line platform.

## Extracted Propagation Metrics

To further evaluate the demonstrated MSSW delay line platform, propagation metrics including group velocity, relaxation time, effective linewidth, unit-time propagation loss, and propagation $Q$-factor ($Q_{PL}$) were extracted from the measured $p$ = 70 μm responses [41]. Fig. 6a compares the extracted group velocity with the theoretical values obtained from the dispersion relation using the transducer selected wavenumber of 448 rad·cm$^{-1}$, corresponding to the 70 μm meander-line pitch, under different applied magnetic fields. The measured trend follows the expected frequency dependence, with the group velocity decreasing from 45 km·s$^{-}$ [1] at 6 GHz to 20 km·s$^{-1}$ at 19.6 GHz. The agreement between the extracted and calculated values confirms that the meander-line transducer selectively excites the intended region of the MSSW dispersion in wavenumber space. Fig. 6b presents the extracted effective ferromagnetic resonance (FMR) linewidth, $\Delta H_{eff}$, and the corresponding effective Gilbert damping, $\alpha_{eff}$, derived from the measured unit-time propagation loss. Following the phenomenological loss treatment described in [31], [40], $\alpha_{eff}$ was obtained from the linear frequency dependence of $\Delta H_{eff}$, yielding a value of $6.92\times10^{-5}$. The detailed conversion procedure is provided in supplementary note 3. This low effective damping is consistent with prior reports showing that interface-related imperfections play a stronger role in magnetic loss when the YIG thickness is reduced to the sub-micrometer regime [40]. Although thinner films can offer lower group velocity and hence higher delay density, practical delay line design requires a trade-off between delay density and loss. The present result therefore supports the use of commercially available high quality, 18 μm thick Liquid Phase Epitaxy (LPE) YIG films as a practical low-loss MSSW delay platform.

For RF delay applications, both absolute insertion loss and unit-time propagation loss are critical performance metrics. Absolute IL determines the link budget of the delay element when integrated into an RF front end, while propagation loss governs how loss accumulates per unit of delay time and therefore sets the maximum achievable delay before signal degradation becomes prohibitive. To capture both aspects, the proposed devices are benchmarked against a comprehensive set of state-of-the-art delay line platforms, including YIG-based spin wave delay lines using backward volume (BV) [32], magnetostatic surface wave (MSSW) [26], [33], [34], [35], and hybrid forward volume-BV [36] modes, surface acoustic wave (SAW) delay lines [5], [42], [43], [44], [45], [46], [47], plate wave acoustic wave delay lines [48], [49], [50], [51], [52], [53], [54], [55], [56], [57], and integrated photonic delay lines [11], [12], [13], [14], [15], [16], [17], [18], [19], [20]. This benchmark is presented progressively across Fig. 6c-f, beginning with absolute IL versus group delay at representative operating frequencies, followed by frequency-resolved unit-time propagation loss and propagation $Q$-factor ($Q_{PL}$).

Fig. 1d, Fig. 6c, and Fig. 6d present the IL versus group delay benchmark at C-band (4-8 GHz), X-band (8-12 GHz), and $K_u$-band (12-18 GHz), where the best-performing devices from each platform within the corresponding frequency range are selected for direct comparison. At C-band, the measured $p$ = 70 μm devices exhibit IL of 2.5 to 3.5 dB across delays of 3.3 to 20.3 ns, substantially lower than YIG MSSW [26], YIG BV

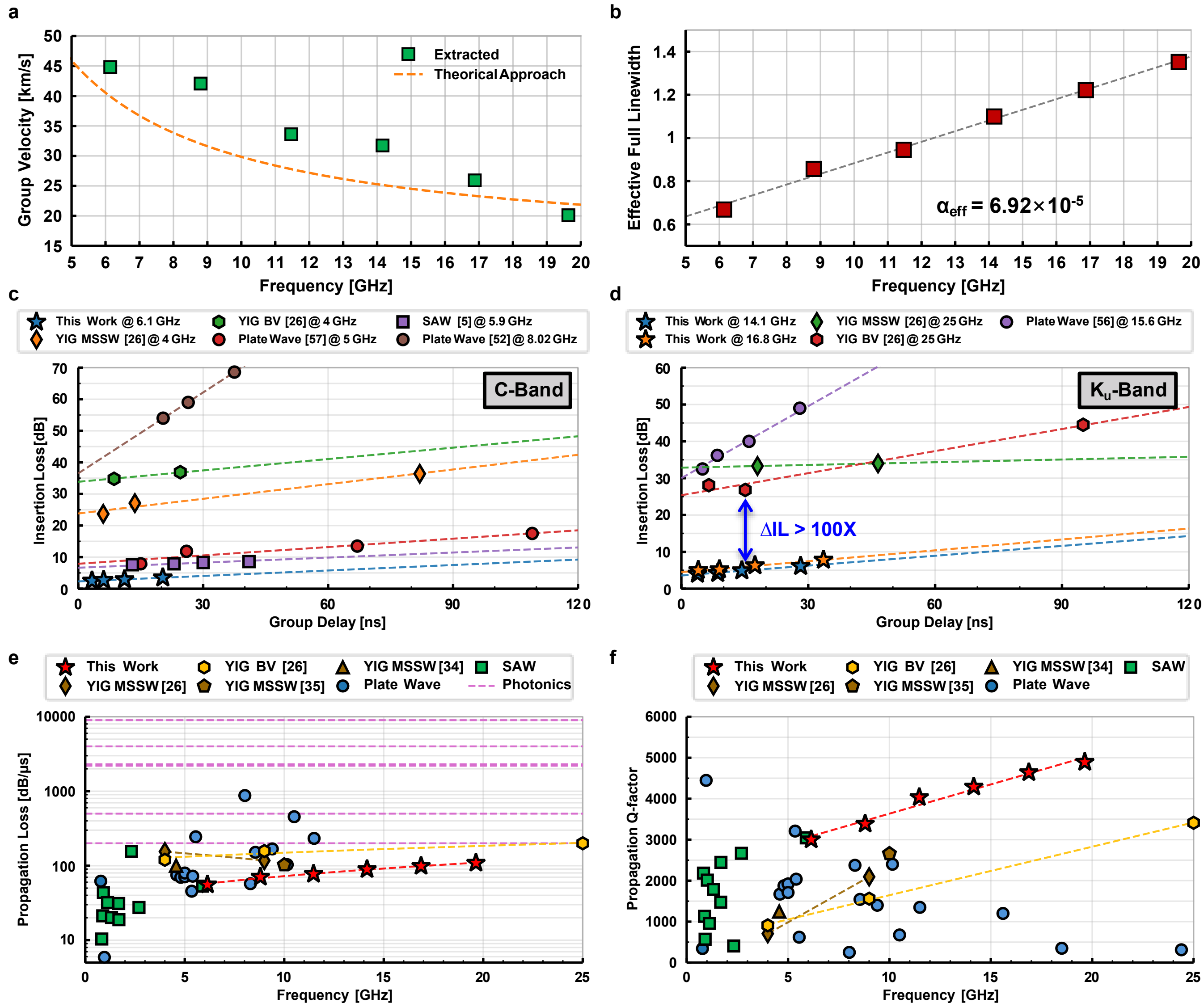

Fig. 6 (a) Extracted group velocity of the $p$ = 70 μm MSSW YIG delay lines compared with the theoretical values calculated from the MSSW dispersion relation. (b) Extracted effective full linewidth as a function of frequency, yielding an effective Gilbert damping of $\alpha_{\text{eff}} = 6.92 \times 10^{-5}$. (c) (d) Insertion loss versus group delay comparison with representative state of the art YIG and acoustic delay lines in the C band (4-8 GHz) and $K_u$ band (12-18 GHz), respectively. (e) Unit-time propagation loss comparison among the measured MSSW YIG delay lines, state-of-the-art spin wave delay lines, acoustic delay lines, and photonic delay lines. (f) Propagation $Q$-factor comparison among the measured MSSW YIG delay line, state-of-the-art spin wave delay lines, and acoustic delay lines.

[26], SAW [5], and plate acoustic wave [52], [57] delay lines operating in a comparable frequency range, as shown in Fig. 6c. Notably, the reported IL of the proposed devices is measured directly under 50 Ω termination without any external impedance matching, whereas the lowest reported IL of inherently fixed frequency acoustic delay lines is typically achieved with dedicated matching networks. This low absolute IL is consistently preserved as the operating frequency is extended to X-band and $K_u$-band. At X-band, the devices exhibit IL of 2.5 to 4.9 dB over delays of 3.3 to 26.4 ns. At comparable delays, the IL is at least 10 dB lower than YIG MSSW [26], YIG BV [26], and plate acoustic wave [53] delay lines reported in the same frequency range, corresponding to more than 10 times lower loss in linear power, as shown in Fig. 1d. At $K_u$-band, the

devices maintain IL of 4 to 7.8 dB over delays of 3.8 to 33.6 ns. At comparable delays, the IL is at least 20 dB lower than YIG MSSW [26], YIG BV [26], and plate acoustic wave [56] platforms reported at similar or lower frequencies, corresponding to over 100 times lower loss in linear power, as shown in Fig. 6d. Across all three benchmarked bands, the slope of IL versus group delay is notably shallow, which directly indicates a low-loss per unit delay accumulation and motivates a frequency domain comparison through a normalized propagation loss metric.

To quantify this observation, the unit-time propagation loss and $Q_{PL}$ are introduced. The unit-time propagation loss reflects the wave attenuation required to accumulate a given amount of delay time while accounting for the group velocity, and $Q_{PL}$ further incorporates the operating frequency to provide a compact measure of wave propagation quality. The $Q_{PL}$ is given in Eq. (5)

$$Q_{PL} = \frac{\pi \times \frac{20}{\ln(10)} \times f_c}{PL} \tag{5}$$

where $f_c$ is the center frequency of the delay line in Hz, $PL$ is the unit-time propagation loss in dB·s$^{-1}$, and the constant is used to convert amplitude between Neper and decibel. As summarized in Fig. 6e, the proposed MSSW delay line exhibits a unit-time propagation loss ranging from 55.7 dB·μs$^{-1}$ at 6 GHz to 109 dB·μs$^{-1}$ at 19.6 GHz. Photonic delay lines, although capable of supporting wide RF bandwidths, generally exhibit unit-time propagation losses several orders of magnitude higher than acoustic and spin wave platforms. Compared with SAW and plate acoustic wave delay lines, the proposed MSSW YIG delay line achieves a unit-time propagation loss comparable to the best-performing acoustic devices below 6 GHz, while remaining extremely low propagation loss into X-band and $K_u$-band. This contrasts with the rapid loss escalation observed in acoustic delay lines at higher frequencies, which reflects the inherent $f{\cdot}Q$ product limitation of acoustic wave propagation [6].

Fig. 6f summarizes $Q_{PL}$ across the same set of platforms, where two clearly distinct trends emerge. The $Q_{PL}$ of acoustic delay lines decreases with frequency, consistent with the $f{\cdot}Q$ product limitation. In contrast, spin wave delay lines show $Q_{PL}$ that increases with frequency, because their unit-time propagation loss grows sub-linearly with frequency. Benefiting from the low propagation loss of the 18 μm thick YIG waveguide, the proposed MSSW delay line exhibits $Q_{PL}$ increasing from 3002 at 6.1 GHz to 4893 at 19.6 GHz, exceeding reported acoustic delay lines across the entire tuning range.

Taken together, the absolute IL benchmarks in Fig. 1d and Fig. 6c and 6d, the unit-time propagation loss in Fig. 6e, and the $Q_{PL}$ trend in Fig. 6f establish the microfabricated YIG MSSW platform as a low-loss, high-frequency delay line technology whose propagation performance remains superior with state-of-the-art acoustic delay lines even at any single operating frequency. On top of this, the platform uniquely provides wideband magnetic field tunability that is inaccessible to acoustic delay lines, enabling true time delay based signal processing in next-generation reconfigurable RF front ends.

# Methods

## Fabrication Process

The reported MSSW YIG delay lines were fabricated based on a commercially available YIG-on-GGG wafer with three lithography steps. The overall fabrication process flow follows our previous work [30]. The 18 μm thick YIG film was grown on the GGG substrate by LPE process. A 500 nm thick $SiO_2$ hard mask was first deposited on the YIG surface and annealed in a nitrogen furnace at 600 °C for 30 min. The hard mask was then patterned by standard photolithography and reactive ion etching, followed by wet etching of the exposed YIG in hot phosphoric acid to define the waveguide geometry. After YIG etching, the remaining $SiO_2$ hard mask was removed in buffered oxide etchant. To reduce the severe topography created by the etched 18 μm thick YIG waveguide, a photosensitive benzocyclobutene (BCB) layer was used for planarization. The BCB was spin coated, patterned to reopen the top surface of the YIG waveguide, and then thermally cured. The transducer antennas and probing pads were then formed from a 2 μm thick aluminum (Al) film deposited by sputtering. A 300 nm thick $SiO_2$ layer was deposited to serve as a hard mask for Al patterning. The $SiO_2$ layer was patterned by photolithography and reactive ion etching. The Al layer was then patterned by inductively coupled plasma etching. Finally, the $SiO_2$ hard mask was removed by reactive ion etching, leaving the patterned Al transducers on the planarized YIG/BCB surface.

## Measurement Setup

The reported MSSW YIG delay lines were measured on a magnetic probe station equipped with electromagnets and a Gaussmeter, enabling accurate control of the applied bias field. RF characterization was carried out using a Keysight P5026B vector network analyzer with 50 Ω port impedance and an input power of -20 dBm. A two port SOLT calibration was performed to the probe tips over the frequency range of interest before measurement. The devices were contacted using 150 μm pitch ground-signal-ground probes. All the data reported in this work did not perform any on wafer de-embedding process.

## Data Availability

All data supporting the findings of this study are available within the article and its supplementary files. Any additional requests for information can be directed to, and will be fulfilled by, the corresponding author.

## Acknowledgement

The authors would like to thank Dr. Todd Bauer, Dr. David Abe and Dr. Tim Hancock of the Defense Advanced Research Projects Agency (DARPA) and Dr. Michael Page of the Air Force Research Laboratory for their guidance and support of this work under the DARPA Wideband Adaptive RF Protection (WARP) program, contract FA8650-21-1-7010. The fabrication of devices was performed at the Singh Center for Nanotechnology, supported by the NSF National Nanotechnology Coordinated Infrastructure Program (No. NNCI-1542153).

## Author contributions

C.-Y. C. and R. O. developed the device concepts and experimental implementations. C.-Y. C., X. D., S. Y., T. W., and S. W. fabricated the YIG delay line under the supervision of R. O., C.-Y. C., X. D., S. Y., T. W., and S. W. performed the filter measurements and R.O. supervised the measurements. C.-Y. C., X. D., S. Y. and R. O. analyzed all data and wrote the manuscript. All authors have given approval to the final version of the manuscript.

# Supplementary Information

## Dispersion Engineered Frequency Tunable Delay Platform based on Magnetostatic Surface Spin Waves

Chin-Yu Chang[1], Xingyu Du[1], Shun Yao[1], Tao Wang[1], Shuxian Wu[1], and Roy H. Olsson III[1]

[1]Department of Electrical and System Engineering, University of Pennsylvania, PA, USA

*Correspondence: Roy H. Olsson III (rolsson@seas.upenn.edu)

# Supplementary Note 1: MSSW propagation band based on dispersion analysis

Owing to the unique dispersion of magnetostatic surface waves (MSSWs), each applied magnetic bias field defines a distinct allowable propagation band. The dispersion relation can be obtained from Eq. (1)-(3) in the manuscript [1], [2]. Here, $k_y$ denotes the wavevector along the MSSW propagation direction, whereas $k_z$is quantized by the finite waveguide width according to $k_{z,n} = n\pi/w$, where $w$ is the waveguide width and $n$ is the width mode order. The resulting MSSW dispersion defines a finite allowable propagation band bounded by a lower cutoff frequency, $f_{\min}$, and an upper cutoff frequency, $f_{\max}$, at each magnetic bias field. SI-Fig. 1-1a shows an example of the calculated dispersion curve and group velocity of the proposed 18 μm YIG waveguide with $w$ = 200 μm at 1500 Gauss. The allowable propagation band is bounded by the $f_{\min}$ and $f_{\max}$. SI-Fig. 1-1b shows the field dependence of this allowable propagation band at each magnetic bias field. As the magnetic field increases, both $f_{\min}$ and $f_{\max}$ shift toward higher frequencies, while the band span, $f_{\max} - f_{\min}$, decreases. Therefore, the allowable MSSW propagation band becomes progressively narrower as the operating frequency shifts upward with increasing magnetic bias field.

To further connect this field-dependent propagation band to device operation, the calculated group velocity is converted into a unit length group delay and mapped to the frequency domain using the MSSW dispersion relation. In parallel, the radiation impedance of the meander line transducer, calculated from Eq. (4) in the main text, is also mapped to the frequency domain using the same dispersion relation. SI-Fig. 1-2 shows the resulting overlays of the unit length group delay and radiation impedance at different magnetic bias fields for two representative transducer pitches, $p$ = 70 μm and $p$ = 210 μm. These two pitches correspond to different dominant excitation wavenumbers and therefore sample different regions of the MSSW dispersion band. The $p$ = 70 μm transducer preferentially excites higher-wavenumber MSSWs near the high-frequency side of the allowable band, where the group velocity is lower and the unit length group delay is larger. In contrast, the $p$ = 210 μm transducer excites lower-wavenumber MSSWs closer to the lower-frequency side of the band, where the unit length group delay is smaller. As the magnetic field increases, the frequency window over which these selected MSSW modes can be efficiently excited becomes narrower, consistent with

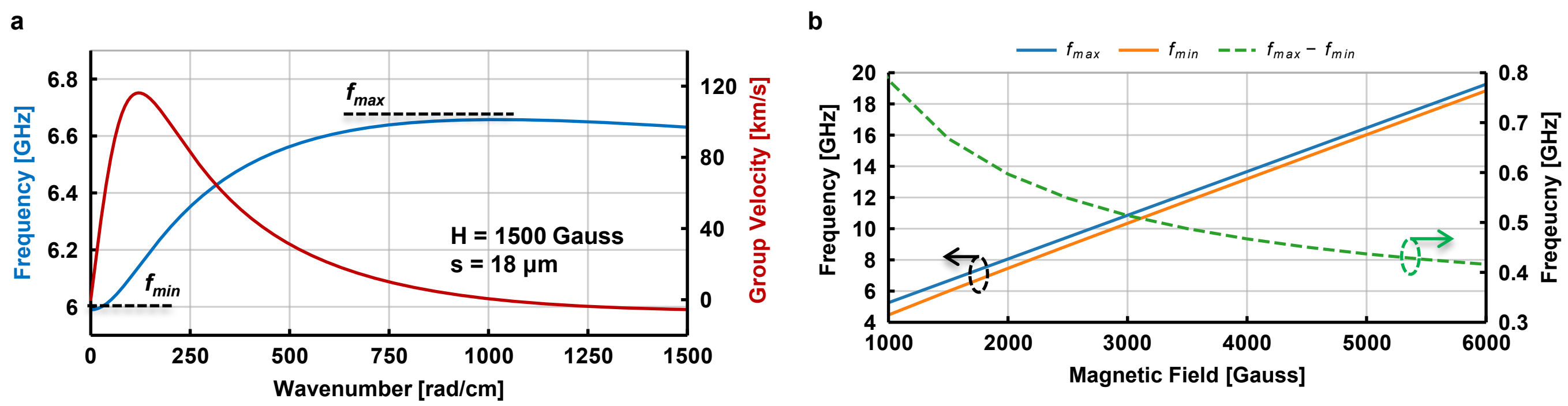

SI-Fig. 1-1 (a) Calculated dispersion curve and group velocity. (b) Calculated allowable propagation band of MSSW in different magnetic bias fields.

the narrowing of the measured transmission response. At the same time, the achievable unit length group delay increases because the calculated group velocity decreases at higher operating frequencies, as discussed in Fig. 2c of the main text. These overlays therefore provide the physical basis for using transducer pitch as a design parameter to tailor the measured MSSW delay response.

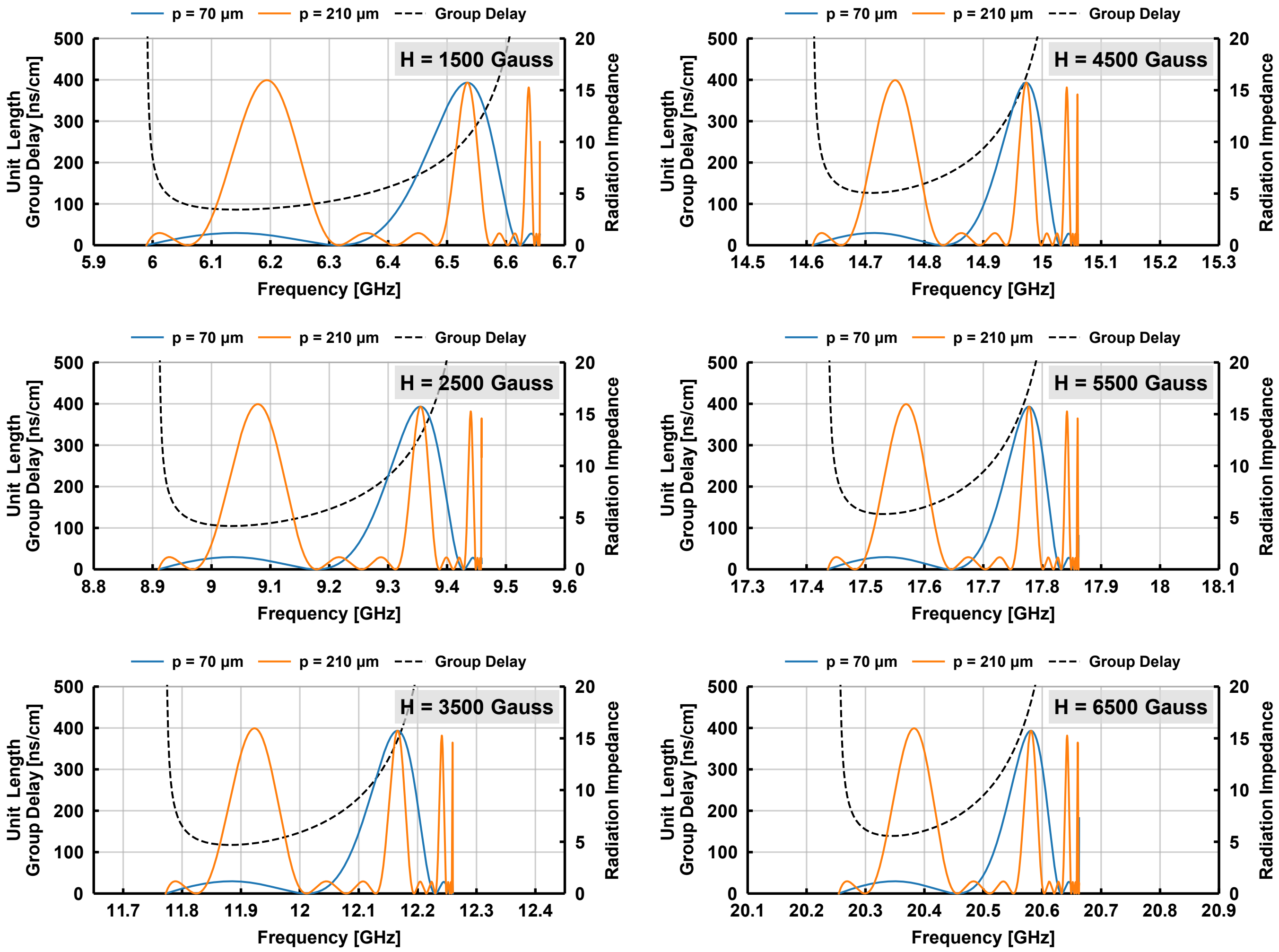

SI-Fig. 1-2 Calculated unit length group delay overlaid with the radiation impedance spectra of meander line transducers with $p$ = 70 μm and $p$ = 210 μm at different magnetic bias fields. The two transducer pitches sample different wavenumber regions of the MSSW dispersion band, resulting in distinct excitation frequencies and group delay values.

# Supplementary Note 2: Zoom-in spectrum and performance summary

This Supplementary Note summarizes the measured performance of the MSSW YIG delay lines with transducer pitches of $p$ = 70 μm and 210 μm. For $p$ = 70 μm devices where $L_{\mathrm{DL}}$ = 70, 210, 420 and 840 μm are included and for the $p$ = 210 μm, devices where $L_{\mathrm{DL}}$ = 210 and 840 μm are presented. Supplementary Figs. 2-1 to 2-12 show the zoom-in transmission spectra, extracted in-band group delay, and performance summaries under applied magnetic fields from 1500 to 6500 G, including center frequency ($f_0$), insertion loss (IL), bandwidth, group delay at the center frequency ($\tau_d$), group velocity ($v_g$), unit-time propagation loss, and propagation $Q$-factor ($Q_{PL}$).

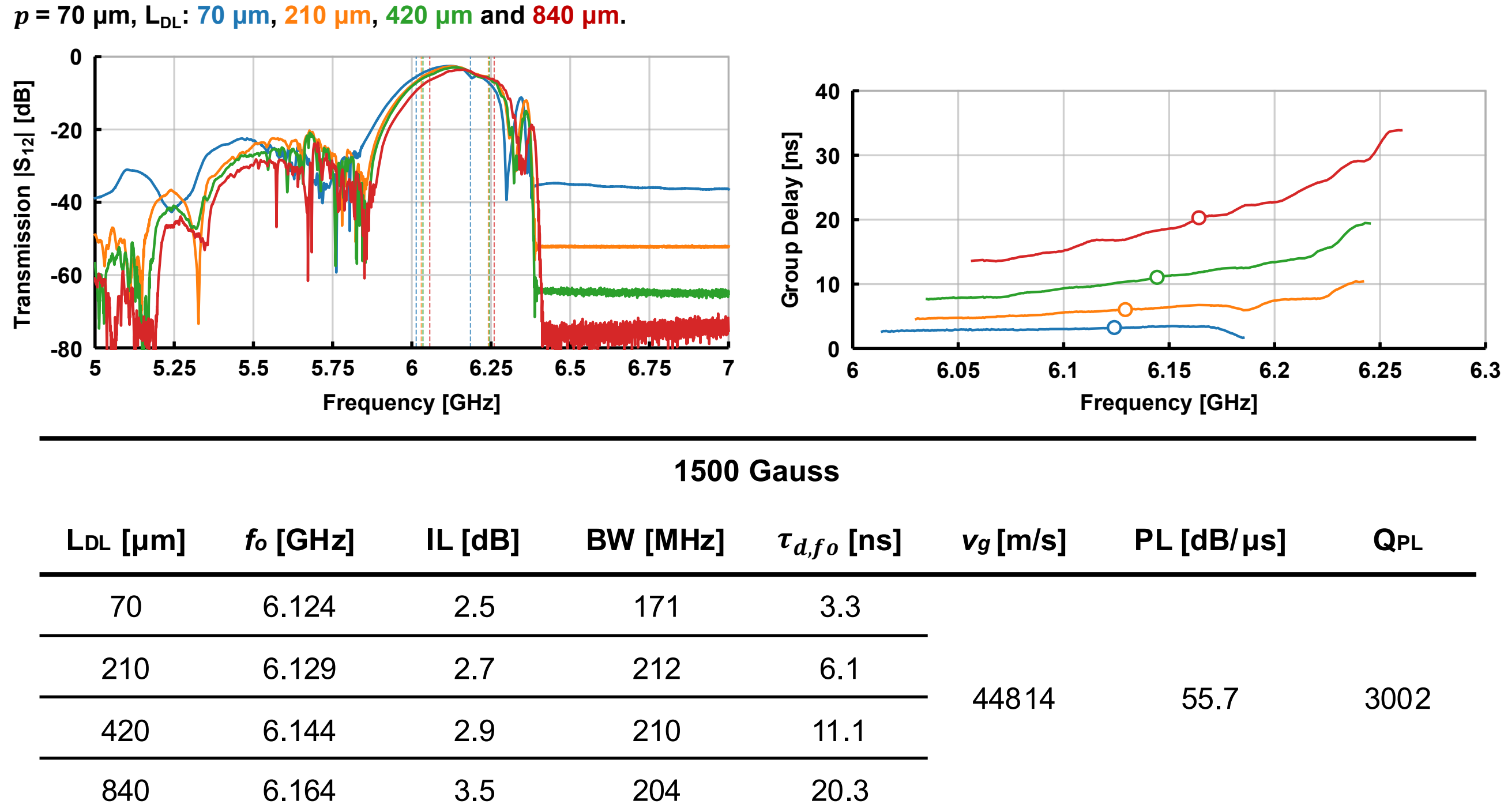

| $L_{DL}$ [μm] | $f_o$ [GHz] | IL [dB] | BW [MHz] | $\tau_{d,fo}$ [ns] | $v_g$ [m/s] | PL [dB/μs] | $Q_{PL}$ |
|---|---|---|---|---|---|---|---|
| 70 | 6.124 | 2.5 | 171 | 3.3 | 44814 | 55.7 | 3002 |
| 210 | 6.129 | 2.7 | 212 | 6.1 | | | |
| 420 | 6.144 | 2.9 | 210 | 11.1 | | | |
| 840 | 6.164 | 3.5 | 204 | 20.3 | | | |

SI-Fig. 2-1 Zoom-in spectrum, extracted in-band group delay, and performance summary of the MSSW YIG delay lines for $p$ = 70 μm at 1500 Gauss.

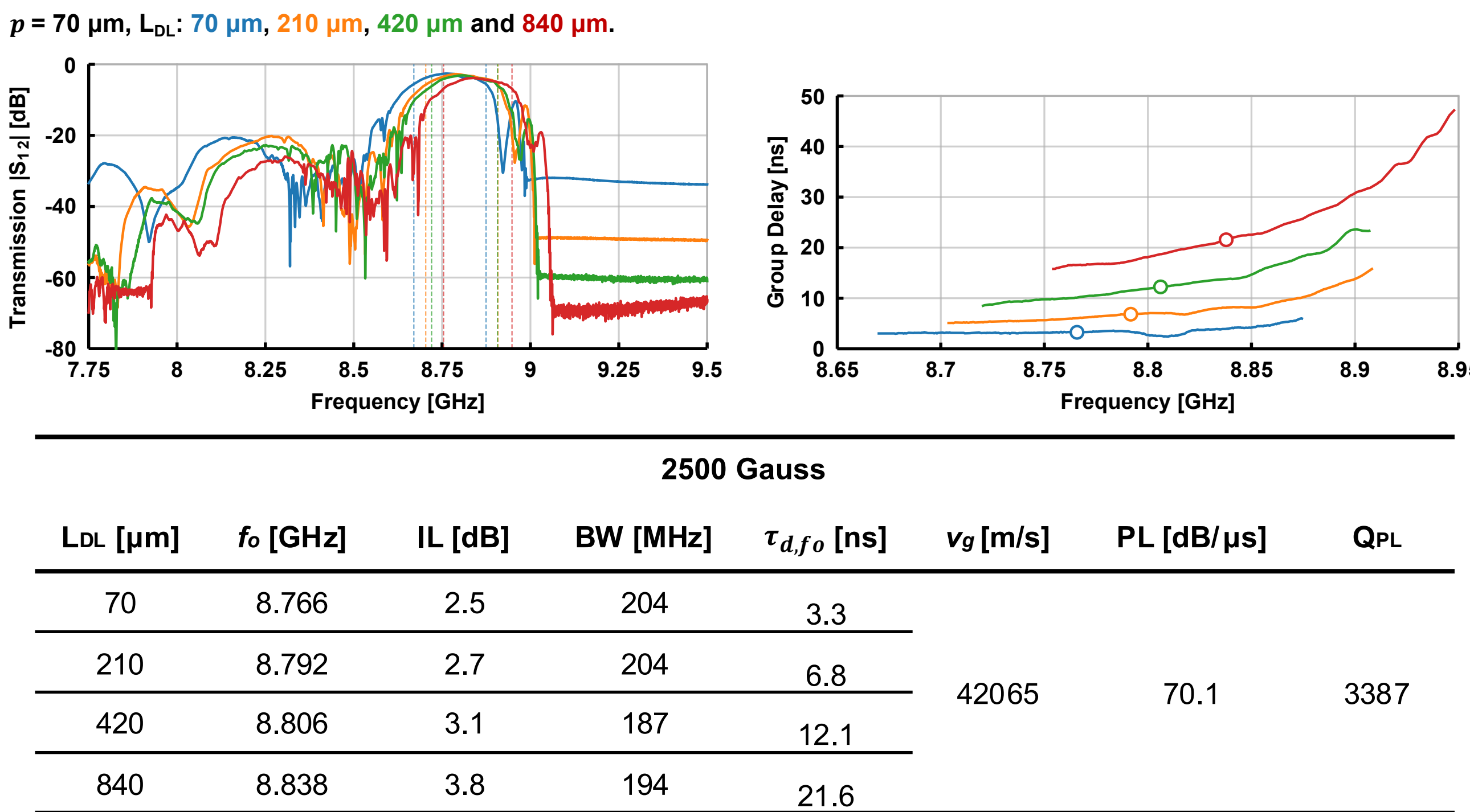

**2500 Gauss**

| $L_{DL}$ [μm] | $f_o$ [GHz] | IL [dB] | BW [MHz] | $\tau_{d,fo}$ [ns] | $v_g$ [m/s] | PL [dB/μs] | $Q_{PL}$ |
|---|---|---|---|---|---|---|---|
| 70 | 8.766 | 2.5 | 204 | 3.3 | 42065 | 70.1 | 3387 |
| 210 | 8.792 | 2.7 | 204 | 6.8 | | | |
| 420 | 8.806 | 3.1 | 187 | 12.1 | | | |
| 840 | 8.838 | 3.8 | 194 | 21.6 | | | |

SI-Fig. 2-2 Zoom-in spectrum, extracted in-band group delay, and performance summary of the MSSW YIG delay lines for $p$ = 70 μm at 2500 Gauss.

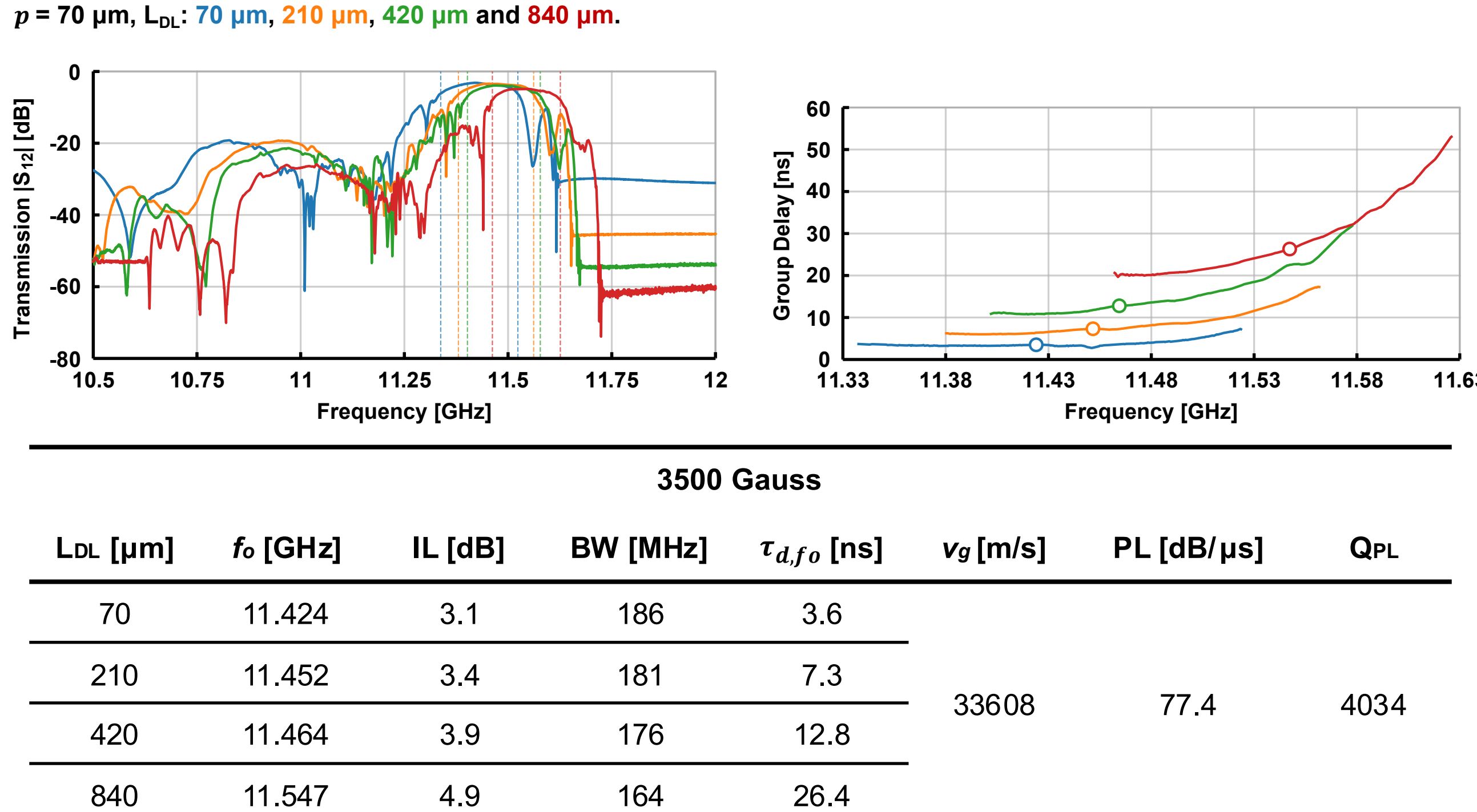

**3500 Gauss**

| $L_{DL}$ [μm] | $f_o$ [GHz] | IL [dB] | BW [MHz] | $\tau_{d,fo}$ [ns] | $v_g$ [m/s] | PL [dB/μs] | $Q_{PL}$ |
|---|---|---|---|---|---|---|---|
| 70 | 11.424 | 3.1 | 186 | 3.6 | 33608 | 77.4 | 4034 |
| 210 | 11.452 | 3.4 | 181 | 7.3 | | | |
| 420 | 11.464 | 3.9 | 176 | 12.8 | | | |
| 840 | 11.547 | 4.9 | 164 | 26.4 | | | |

SI-Fig. 2-3 Zoom-in spectrum, extracted in-band group delay, and performance summary of the MSSW YIG delay lines for $p$ = 70 μm at 3500 Gauss.

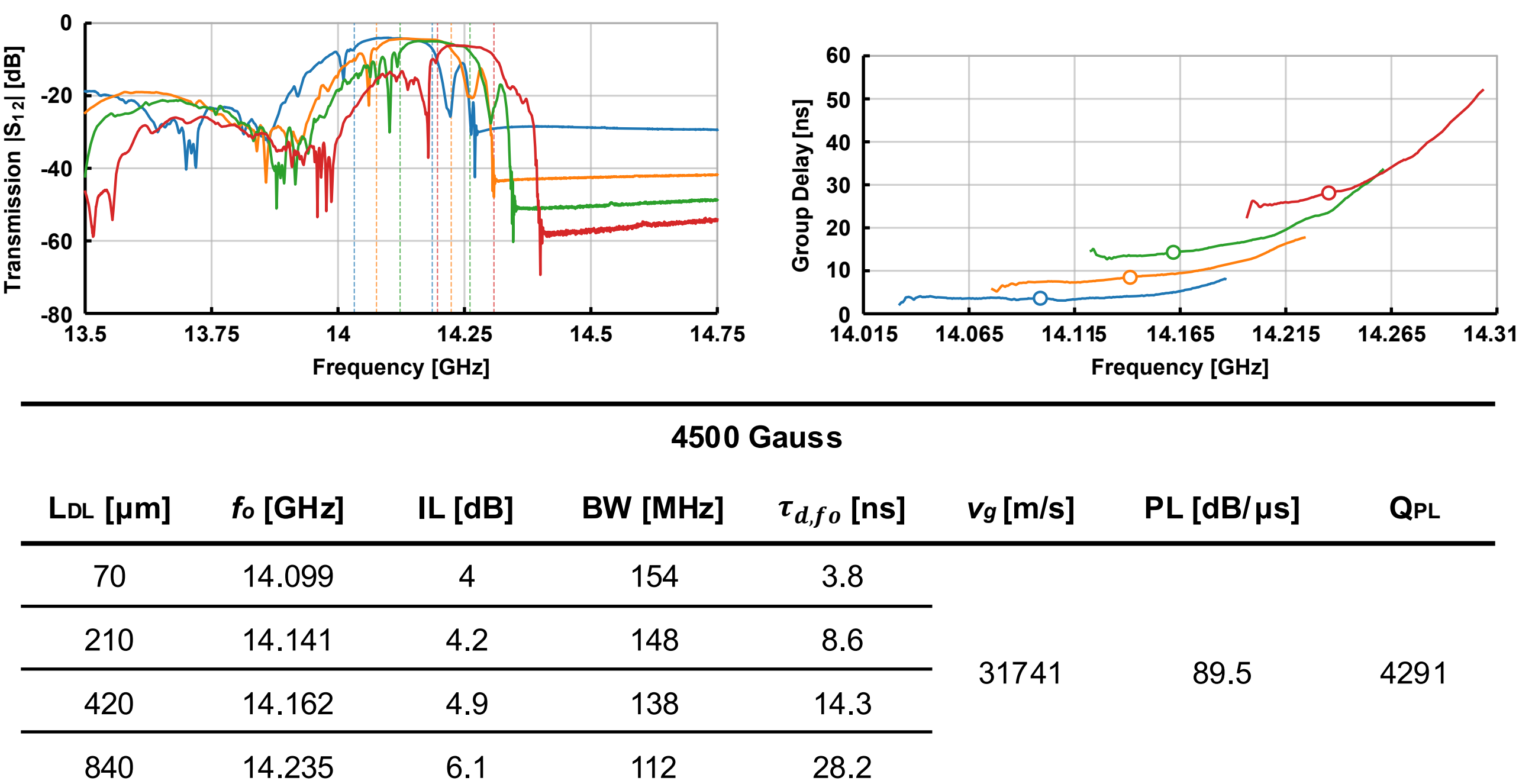

| 4500 Gauss | | | | | | | |
|---|---|---|---|---|---|---|---|
| $L_{DL}$ [μm] | $f_o$ [GHz] | IL [dB] | BW [MHz] | $\tau_{d,fo}$ [ns] | $v_g$ [m/s] | PL [dB/μs] | $Q_{PL}$ |
| 70 | 14.099 | 4 | 154 | 3.8 | 31741 | 89.5 | 4291 |
| 210 | 14.141 | 4.2 | 148 | 8.6 | | | |
| 420 | 14.162 | 4.9 | 138 | 14.3 | | | |
| 840 | 14.235 | 6.1 | 112 | 28.2 | | | |

SI-Fig. 2-4 Zoom-in spectrum, extracted in-band group delay, and performance summary of the MSSW YIG delay lines for *p* = 70 μm at 4500 Gauss.

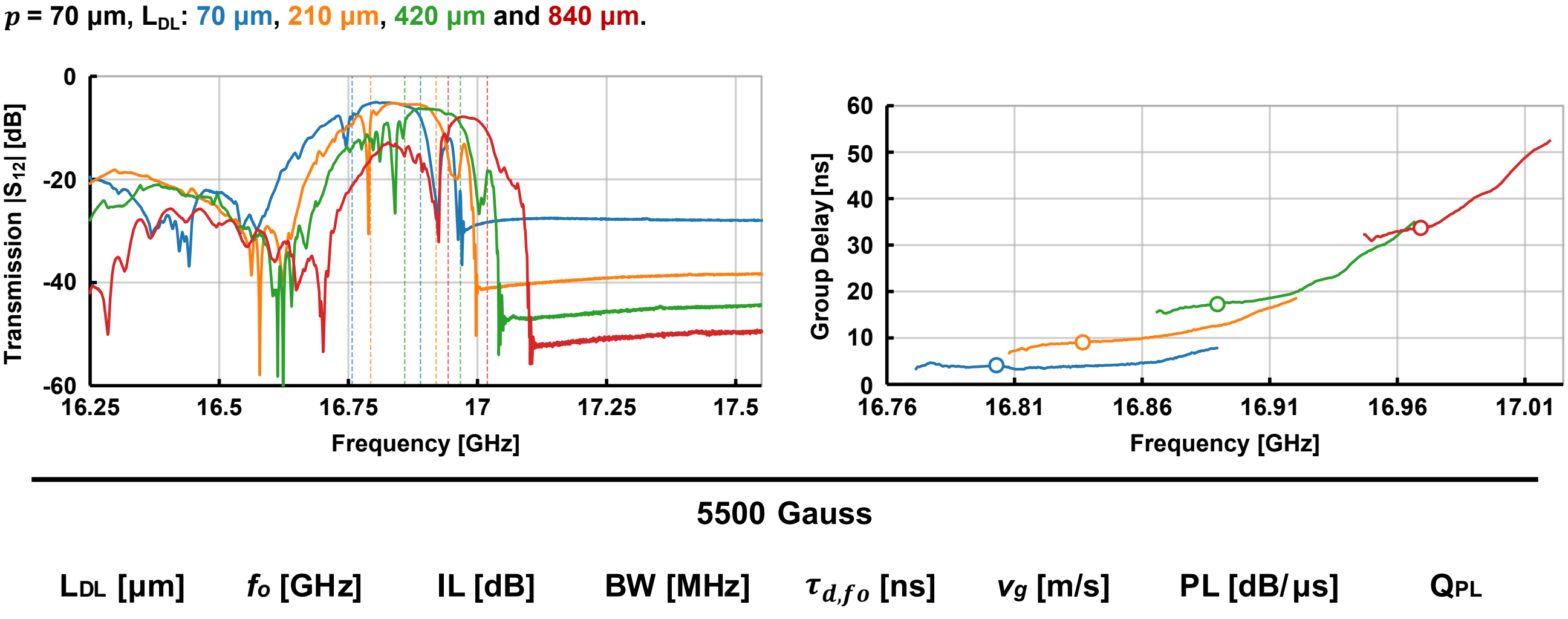

| 5500 Gauss | | | | | | | |
|---|---|---|---|---|---|---|---|
| $L_{DL}$ [μm] | $f_o$ [GHz] | IL [dB] | BW [MHz] | $\tau_{d,fo}$ [ns] | $v_g$ [m/s] | PL [dB/μs] | $Q_{PL}$ |
| 70 | 16.803 | 5 | 132 | 4 | 25923 | 99 | 4641 |
| 210 | 16.837 | 5.2 | 127 | 9 | | | |
| 420 | 16.89 | 6.3 | 108 | 17.4 | | | |
| 840 | 16.969 | 7.8 | 76 | 33.6 | | | |

SI-Fig. 2-5 Zoom-in spectrum, extracted in-band group delay, and performance summary of the MSSW YIG delay lines for *p* = 70 μm at 5500 Gauss.

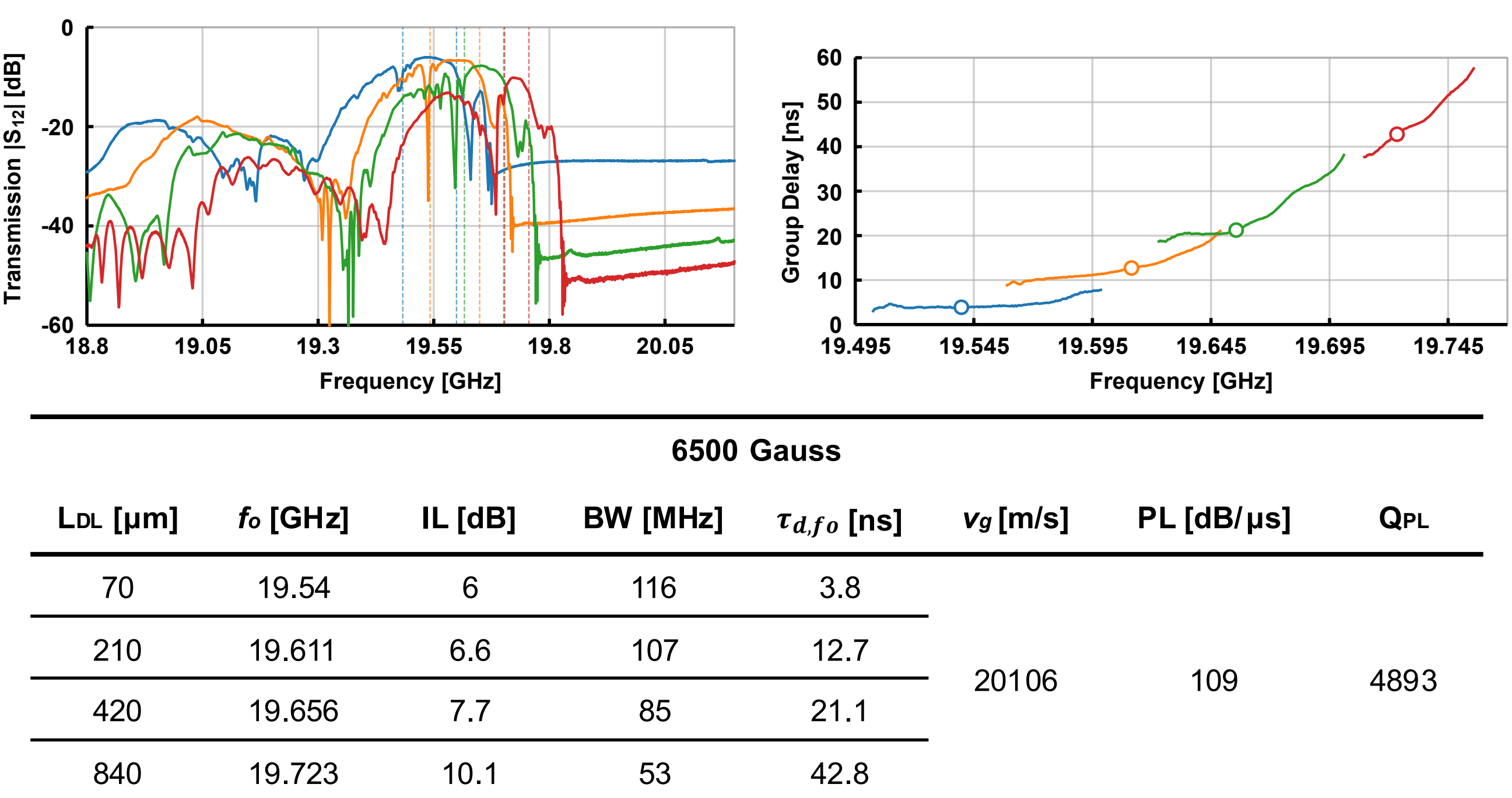

**6500 Gauss**

| $L_{DL}$ [μm] | $f_o$ [GHz] | IL [dB] | BW [MHz] | $\tau_{d,fo}$ [ns] | $v_g$ [m/s] | PL [dB/μs] | $Q_{PL}$ |
|---|---|---|---|---|---|---|---|
| 70 | 19.54 | 6 | 116 | 3.8 | 20106 | 109 | 4893 |
| 210 | 19.611 | 6.6 | 107 | 12.7 | | | |
| 420 | 19.656 | 7.7 | 85 | 21.1 | | | |
| 840 | 19.723 | 10.1 | 53 | 42.8 | | | |

SI-Fig. 2-6 Zoom-in spectrum, extracted in-band group delay, and performance summary of the MSSW YIG delay lines for $p$ = 70 μm at 6500 Gauss.

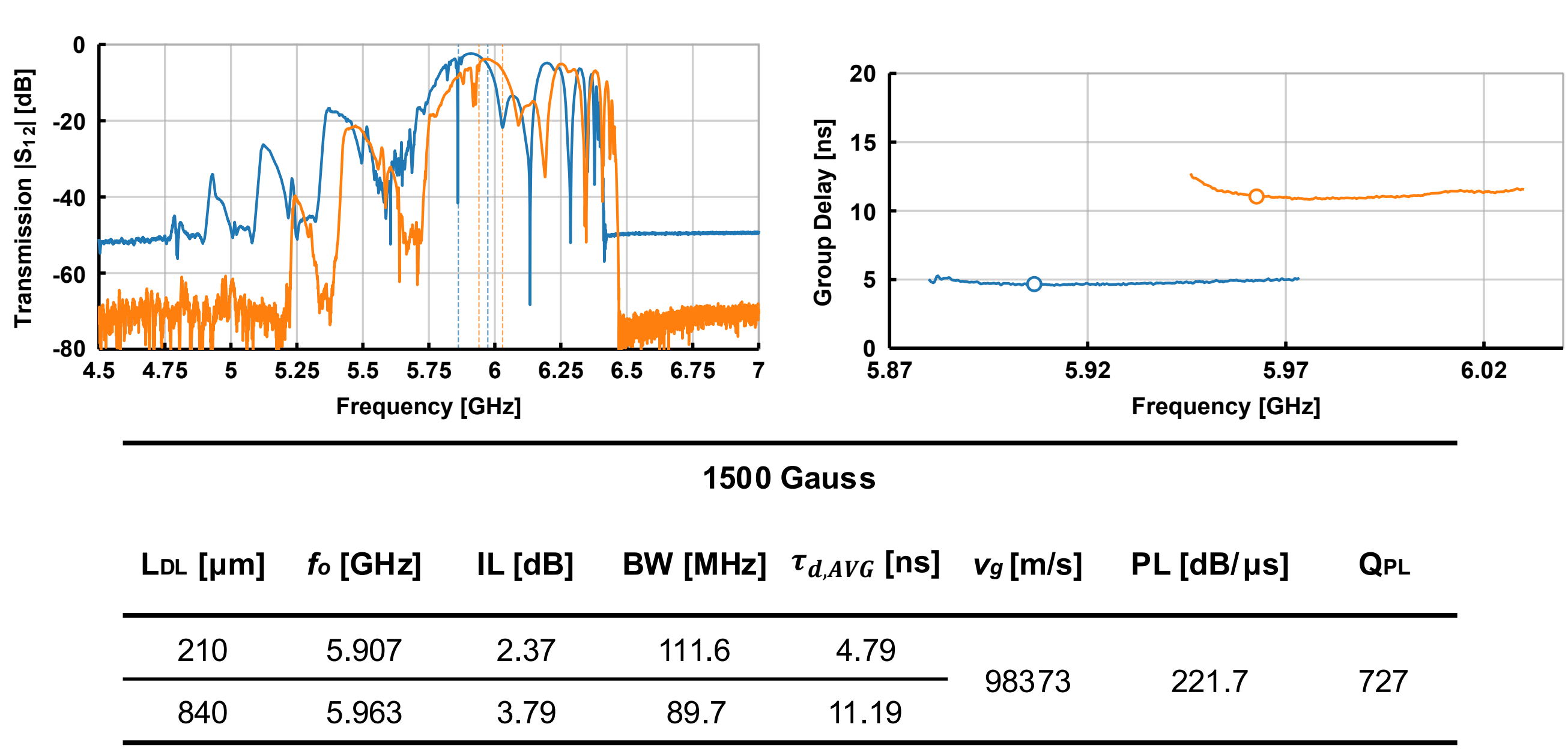

**1500 Gauss**

| $L_{DL}$ [μm] | $f_o$ [GHz] | IL [dB] | BW [MHz] | $\tau_{d,AVG}$ [ns] | $v_g$ [m/s] | PL [dB/μs] | $Q_{PL}$ |
|---|---|---|---|---|---|---|---|
| 210 | 5.907 | 2.37 | 111.6 | 4.79 | 98373 | 221.7 | 727 |
| 840 | 5.963 | 3.79 | 89.7 | 11.19 | | | |

SI-Fig. 2-7 Zoom-in spectrum, extracted in-band group delay, and performance summary of the MSSW YIG delay lines for $p$ = 210 μm at 1500 Gauss.

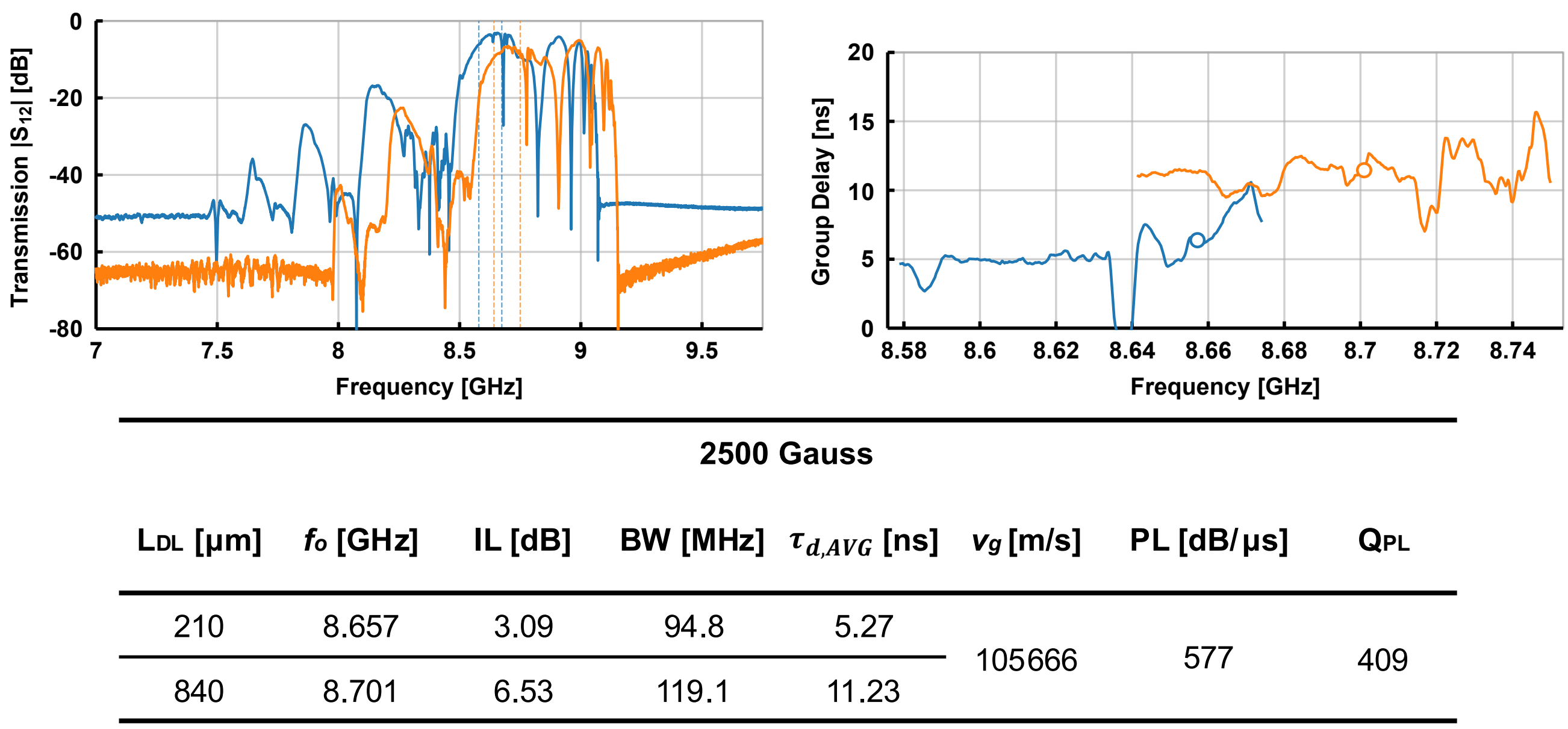

**2500 Gauss**

| $L_{DL}$ [µm] | $f_o$ [GHz] | IL [dB] | BW [MHz] | $\tau_{d,AVG}$ [ns] | $v_g$ [m/s] | PL [dB/µs] | $Q_{PL}$ |
|---|---|---|---|---|---|---|---|
| 210 | 8.657 | 3.09 | 94.8 | 5.27 | 105666 | 577 | 409 |
| 840 | 8.701 | 6.53 | 119.1 | 11.23 | | | |

SI-Fig. 2-8 Zoom-in spectrum, extracted in-band group delay, and performance summary of the MSSW YIG delay lines for $p$ = 210 µm at 2500 Gauss.

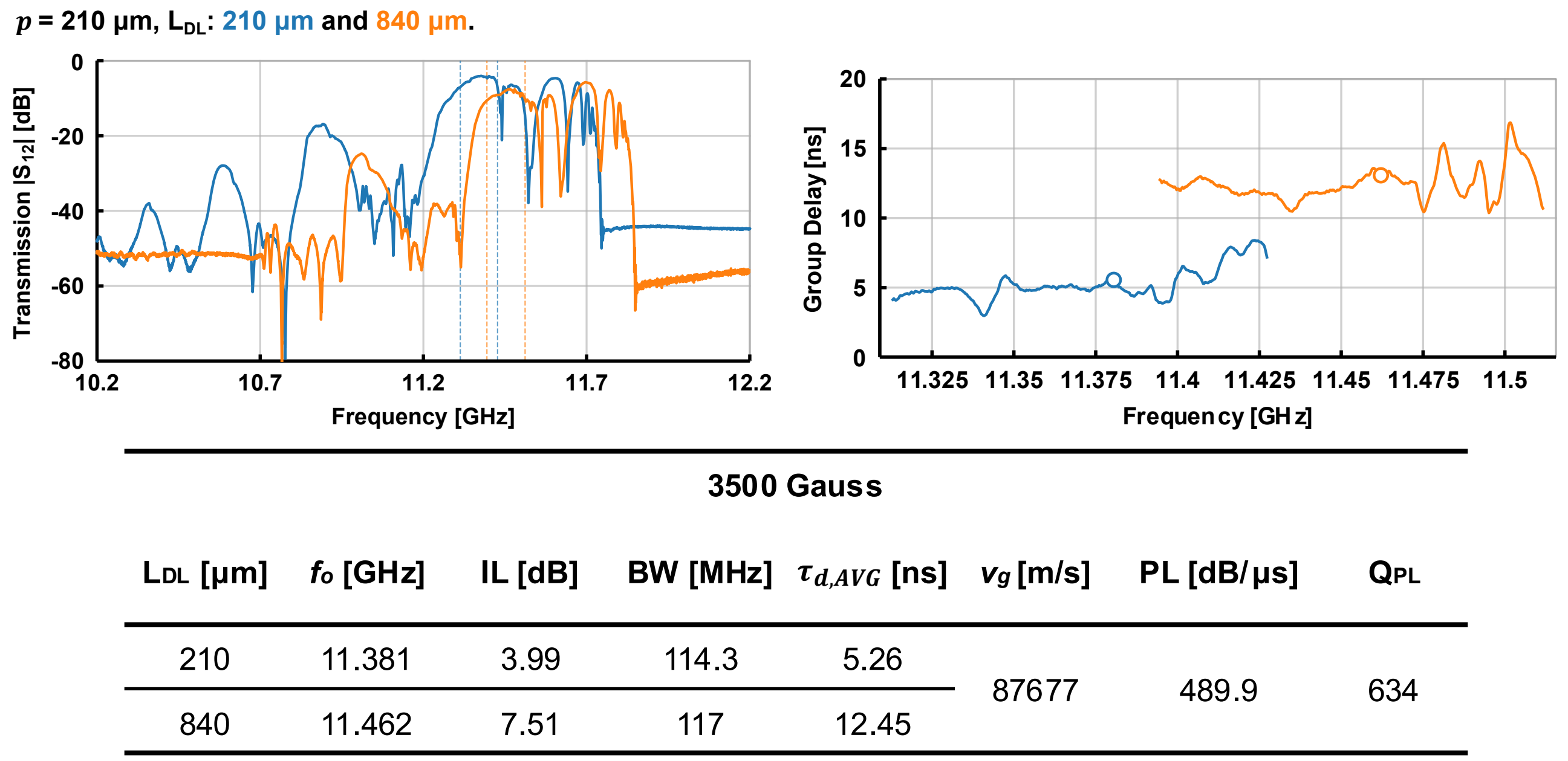

**3500 Gauss**

| $L_{DL}$ [µm] | $f_o$ [GHz] | IL [dB] | BW [MHz] | $\tau_{d,AVG}$ [ns] | $v_g$ [m/s] | PL [dB/µs] | $Q_{PL}$ |
|---|---|---|---|---|---|---|---|
| 210 | 11.381 | 3.99 | 114.3 | 5.26 | 87677 | 489.9 | 634 |
| 840 | 11.462 | 7.51 | 117 | 12.45 | | | |

SI-Fig. 2-9 Zoom-in spectrum, extracted in-band group delay, and performance summary of the MSSW YIG delay lines for $p$ = 210 µm at 3500 Gauss.

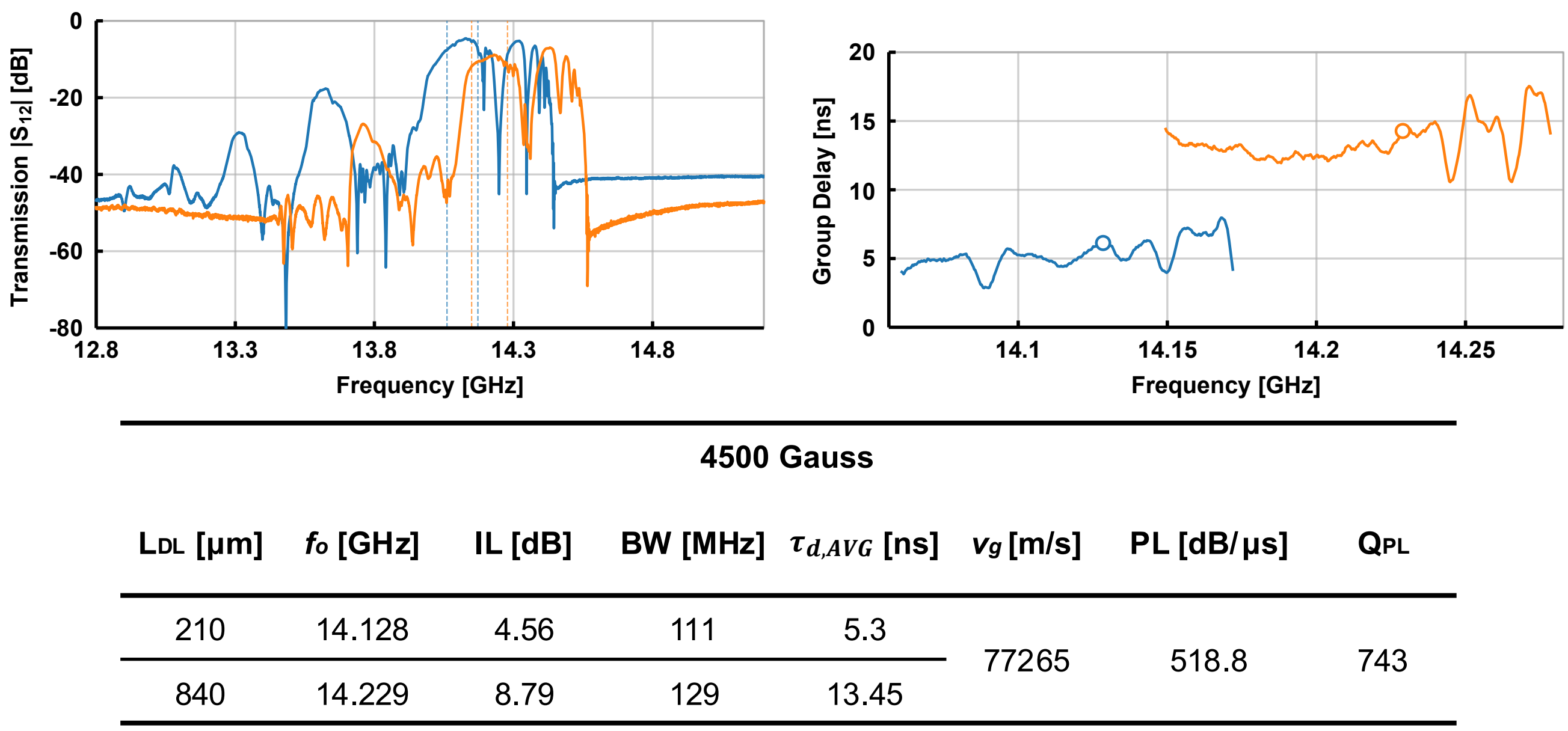

**4500 Gauss**

| $L_{DL}$ [μm] | $f_o$ [GHz] | IL [dB] | BW [MHz] | $\tau_{d,AVG}$ [ns] | $v_g$ [m/s] | PL [dB/μs] | $Q_{PL}$ |
|---|---|---|---|---|---|---|---|
| 210 | 14.128 | 4.56 | 111 | 5.3 | 77265 | 518.8 | 743 |
| 840 | 14.229 | 8.79 | 129 | 13.45 | | | |

SI-Fig. 2-10 Zoom-in spectrum, extracted in-band group delay, and performance summary of the MSSW YIG delay lines for $p$ = 210 μm at 4500 Gauss.

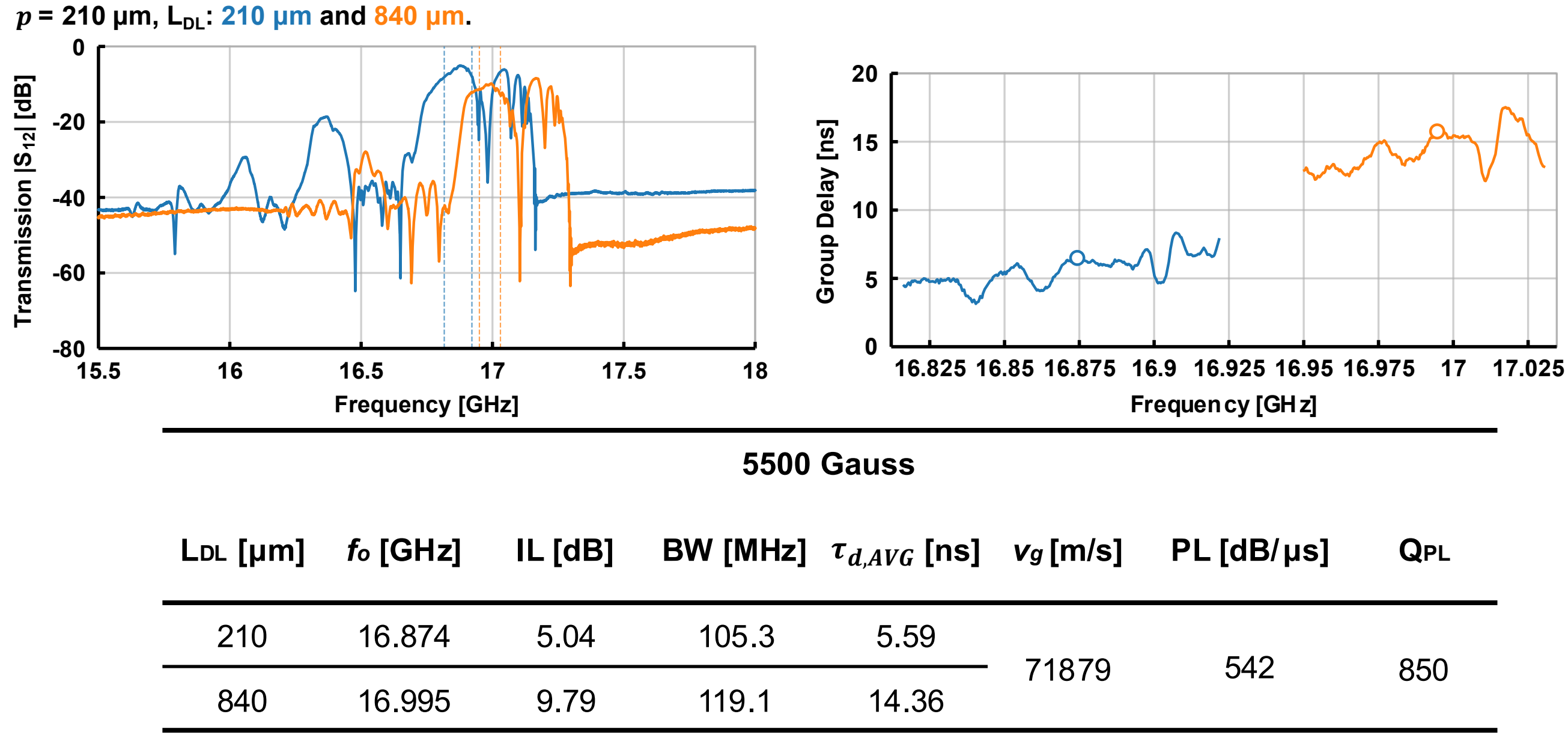

**5500 Gauss**

| $L_{DL}$ [μm] | $f_o$ [GHz] | IL [dB] | BW [MHz] | $\tau_{d,AVG}$ [ns] | $v_g$ [m/s] | PL [dB/μs] | $Q_{PL}$ |
|---|---|---|---|---|---|---|---|
| 210 | 16.874 | 5.04 | 105.3 | 5.59 | 71879 | 542 | 850 |
| 840 | 16.995 | 9.79 | 119.1 | 14.36 | | | |

SI-Fig. 2-11 Zoom-in spectrum, extracted in-band group delay, and performance summary of the MSSW YIG delay lines for $p$ = 210 μm at 5500 Gauss.

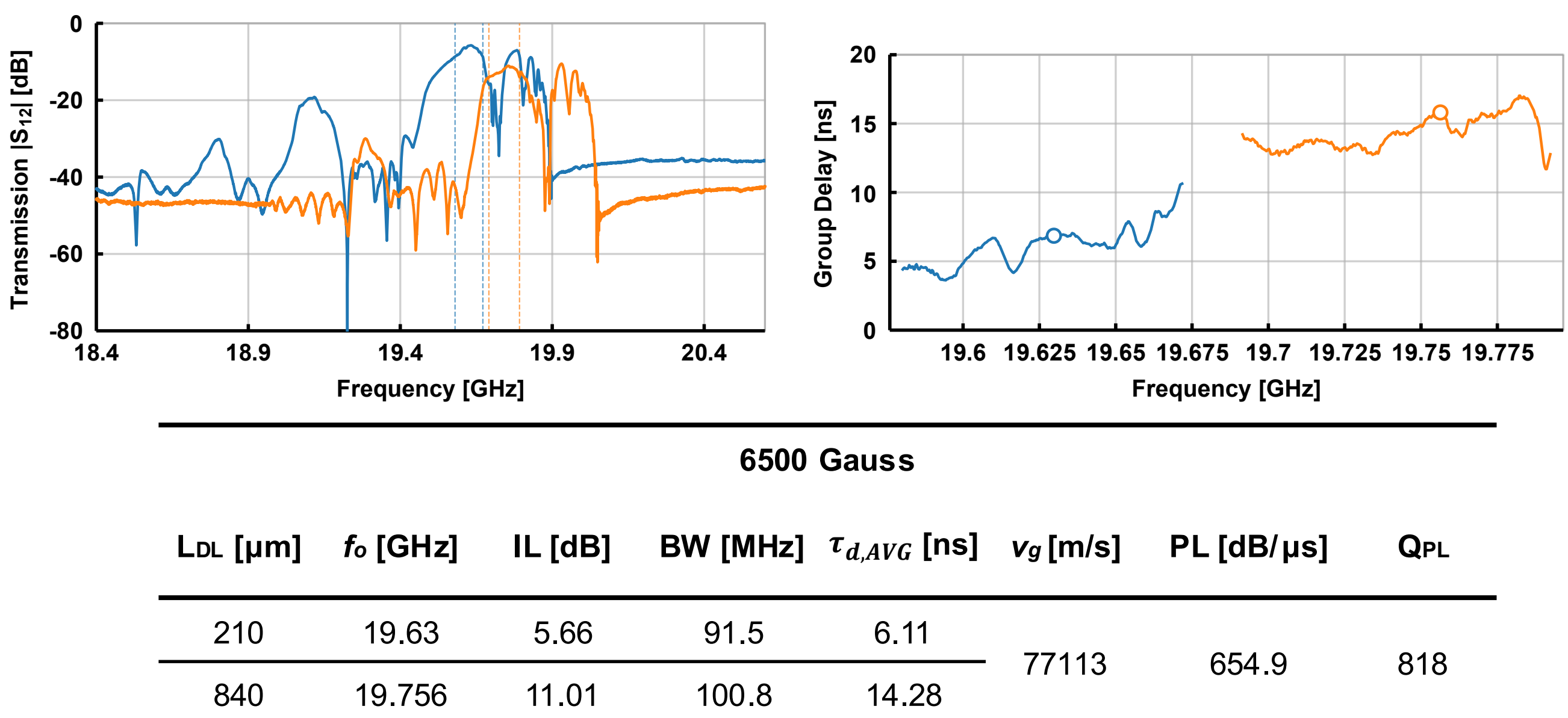

**6500 Gauss**

| $L_{DL}$ [μm] | $f_o$ [GHz] | IL [dB] | BW [MHz] | $\tau_{d,AVG}$ [ns] | $v_g$ [m/s] | PL [dB/μs] | $Q_{PL}$ |
|---|---|---|---|---|---|---|---|
| 210 | 19.63 | 5.66 | 91.5 | 6.11 | 77113 | 654.9 | 818 |
| 840 | 19.756 | 11.01 | 100.8 | 14.28 | | | |

SI-Fig. 2-12 Zoom-in spectrum, extracted in-band group delay, and performance summary of the MSSW YIG delay lines for $p$ = 210 μm at 6500 Gauss.

# Supplementary Note 3: Effective Gilbert damping extraction

To further quantify energy dissipation in the proposed MSSW platform, the measured unit-time propagation loss is converted into an effective relaxation time, an effective linewidth, and finally an effective Gilbert damping. Following the phenomenological propagation loss treatment [3], [4], the measured unit-time propagation loss $PL$ is first converted into an effective modal relaxation time $T_{k,\mathrm{eff}}$ using

$$PL = \frac{8.686}{T_{k,eff}} \times 10^{-6} \tag{1}$$

where $PL$ is expressed in dB·μs$^{-1}$. The effective linewidth is then obtained from

$$\frac{1}{T_{k,eff}} = \frac{|\gamma| \cdot \Delta H_{eff}}{2} \tag{2}$$

The effective Gilbert damping is subsequently extracted from the linear frequency dependence of $\Delta H_{\mathrm{eff}}$. Following the standard linewidth fitting procedure, $\Delta H_{\mathrm{eff}}$ is fitted as

$$\Delta H_{eff} = \Delta H_{0,eff} + \frac{2\alpha_{eff}}{\gamma} \cdot f \tag{3}$$

where $\Delta H_{0,\mathrm{eff}}$ is the zero-frequency intercept, $\gamma$ is the gyromagnetic ratio in Hz/Oe, and $f$ is the operational frequency. Therefore, the effective Gilbert damping is obtained from the slope, $m$, through

$$\alpha_{eff} = \frac{\gamma}{2} m \tag{4}$$

In this work, $\gamma = 2.8$ MHz/G is used for the conversion. Here, $\Delta H_{\mathrm{eff}}$ and $\alpha_{\mathrm{eff}}$ are referred to as effective quantities because they are inferred from the measured propagation loss through the phenomenological loss model rather than measured directly from FMR spectra. Using this procedure, the effective linewidth and effective Gilbert damping shown in Fig. 6b of the manuscript are obtained from the measured unit-time propagation loss.

# Supplementary Note 4: Benchmarking with photonic delay lines

This note compiles the source data used for the benchmarking results shown in Fig. 6c. The table summarizes the unit-time propagation loss and achievable maximum delay of photonic delay lines reported on different material platforms.

SI Table 4-1: Performance summary of the state-of-the-art photonics delay lines.

| Ref | PL [dB/μs] | Maximum Delay [ns] | Platform |
|---|---|---|---|
| [5] | 200 | 12.35 | SiN |
| [6] | 500 | 1.163 | SOI |
| [7] | 2200 | 0.3955 | SiN |
| [8] | 2300 | 5.11 | SOI |
| [9] | 4000 | 12.7 | SOI |
| [10] | 9000 | 1.27 | SOI |
| [11] | 19700 | 0.19137 | SOI |
| [12] | 23000 | 0.15 | Thin-Film LN |
| [13] | 26000 | 0.0902 | $SIO_2$ |
| [14] | 73000 | 1.27 | SOI |
| This work @ 6.13 GHz | 55.68 | 20.3 | MSSW YIG |
| This work @ 8.8 GHz | 70.60 | 21.6 | |
| This work @ 11.45 GHz | 77.40 | 26.4 | |
| This work @ 14.1 GHz | 89.54 | 28.2 | |
| This work @ 16.82 GHz | 98.95 | 33.6 | |
| This work @ 19.6 GHz | 108.87 | 42.8 | |

# Supplementary Note 5: Benchmarking with acoustic delay lines.

This note provides the detailed benchmarking data for the acoustic delay lines used to generate Fig. 6c and d. It summarizes the center frequency, unit-time propagation loss, and propagation *Q*-factor of surface acoustic wave (SAW) and plate wave delay lines implemented on different piezoelectric material platforms.

SI Table 5-1: Performance summary of the state-of-the-art acoustic delay lines.

| Ref | Frequency [GHz] | PL [dB/μs] | Propagation Q-factor | Technology | Platform |
|---|---|---|---|---|---|
| [15] | 0.78 | 62 | 343 | Plate Wave | AlScN |
| [16] | 0.83 | 10.4 | 2184 | SAW | LN/$SiO_2$/Sapphire |
| [16] | 0.88 | 21.2 | 1130 | SAW | LN/$SiO_2$/Si |
| [17] | 0.91 | 43.5 | 571 | SAW | LN/$SiO_2$/Si |
| [18] | 0.96 | 5.9 | 4447 | Plate Wave | LN |
| [19] | 1.02 | -- | 2013 | SAW | LN-on-SiC |
| [20] | 1.12 | 32 | 957 | SAW | LN-on-Sapp |
| [21] | 1.32 | 20.2 | 1787 | SAW | LN-on-SiC |
| [22] | 1.68 | 31.1 | 1475 | SAW | LT/$SiO_2$/Si |
| [22] | 1.69 | 18.9 | 2446 | SAW | LT/$SiO_2$/Si |
| [21] | 2.32 | 155.9 | 406 | SAW | LN-on-SiC |
| [19] | 2.70 | 27.5 | 2668 | SAW | LN-on-SiC |
| [23] | 4.60 | 75.1 | 1671 | Plate Wave | LN |
| [23] | 4.80 | 69.8 | 1876 | Plate Wave | LN |
| [23] | 5.00 | 71 | 1921 | Plate Wave | LN |
| [23] | 5.00 | 79.7 | 1711 | Plate Wave | LN |
| [23] | 5.35 | 45.5 | 3210 | Plate Wave | LN |
| [24] | 5.40 | 72.4 | 2036 | Plate Wave | LN |
| [25] | 5.55 | 244.5 | 620 | Plate Wave | AlScN |
| [26] | 5.90 | 53.2 | 3044 | SAW | AlScN-on-Sapp |
| [25] | 8.02 | 875.4 | 250 | Plate Wave | AlScN |
| [27] | 8.30 | 57.3 | 2375 | Plate Wave | AlScN |
| [28] | 8.55 | 151.2 | 1543 | Plate Wave | LN |
| [27] | 9.40 | 167.9 | 1400 | Plate Wave | AlScN |
| [27] | 10.15 | 103.3 | 2400 | Plate Wave | AlScN |
| [29] | 10.50 | 454.7 | 675 | Plate Wave | AlScN |
| [28] | 11.50 | 232.8 | 1348 | Plate Wave | LN |
| [30] | 15.60 | -- | 1200 | Plate Wave | 2-layer P3F LN |
| [29] | 18.50 | -- | 350 | Plate Wave | AlScN |
| [29] | 24.40 | -- | 313 | Plate Wave | AlScN |
| This Work | 6.14 | 55.7 | 3002 | MSSW | YIG-on-GGG |
| | 8.8 | 70.6 | 3387 | | |
| | 11.47 | 77.4 | 4034 | | |
| | 14.16 | 89.5 | 4291 | | |
| | 16.87 | 99 | 4641 | | |
| | 19.63 | 108.9 | 4893 | | |

# Supplementary Note 6: Graphical fabrication process flow and fabrication results

The detailed fabrication process is described in the Methods section of the manuscript. To provide a clearer overview of the device fabrication sequence, this supplementary note includes a graphical process flow, as shown in SI-Fig. 6-1. Optical images of the four devices with different waveguide lengths reported in this work are also provided in SI-Fig. 6-2.

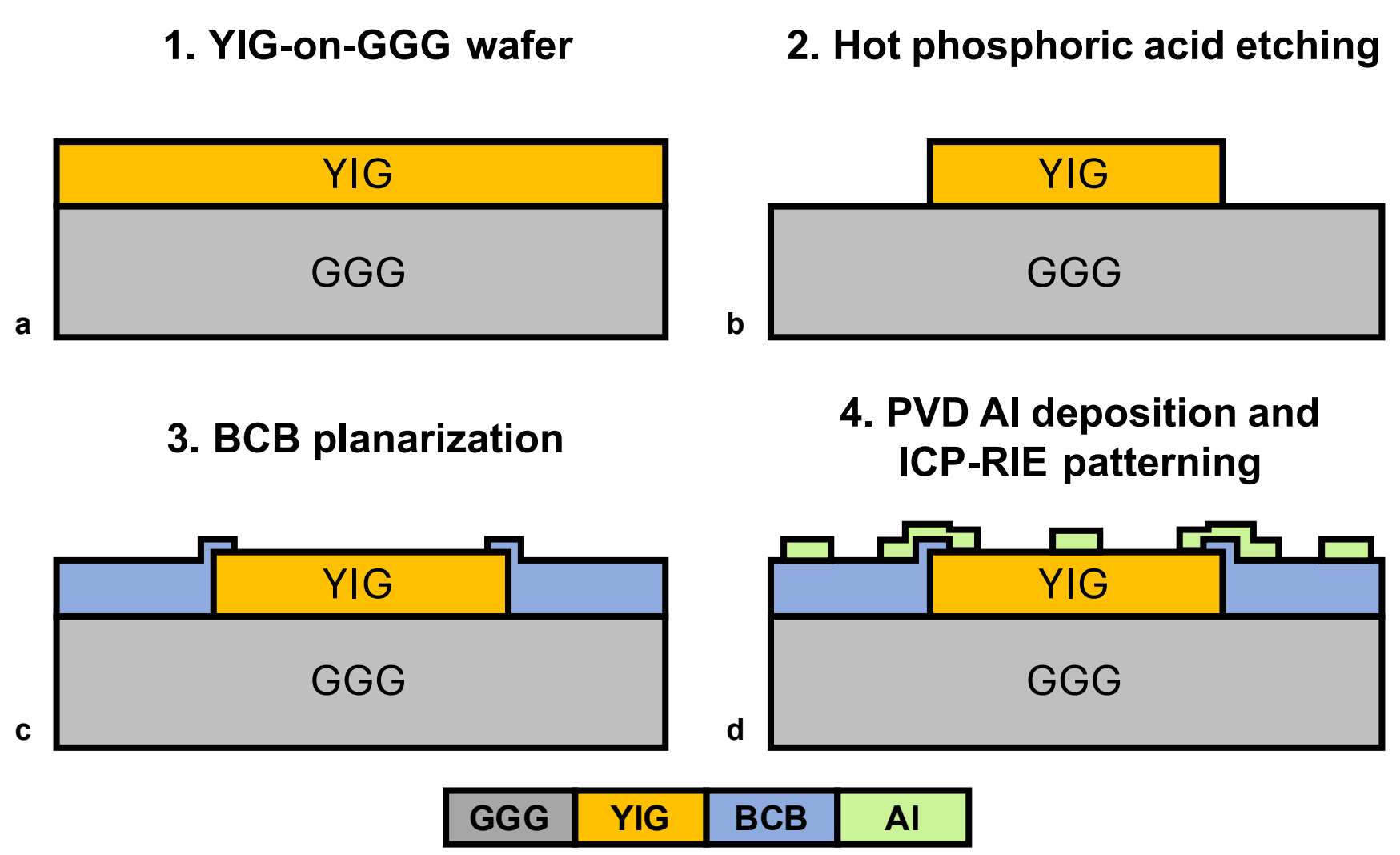

SI-Fig. 6-1 Graphical fabrication process flow used in this work.

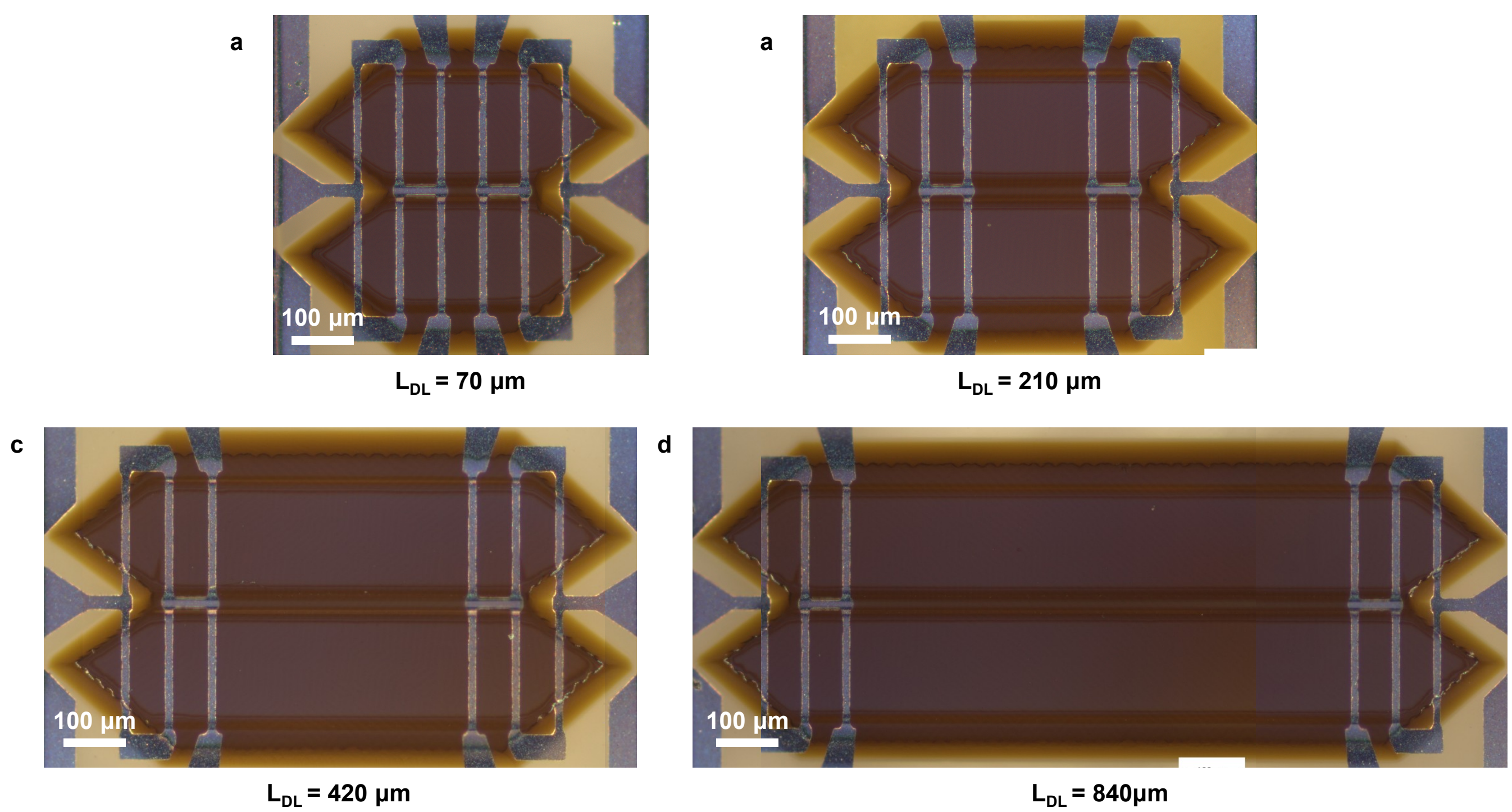

SI-Fig. 6-2 Optical image of the fabricated device with waveguide length (LDL) of (a) 70 μm, (b) 210 μm, (c) 420 μm, and (d) 840 μm.

10.1109/IMS37964.2023.10187967.